\documentclass{emulateapj}

\usepackage{natbib}
\usepackage{amsmath}
\usepackage{graphicx}
\usepackage{epstopdf}

\newcommand{\Msun}{M$_{\odot}$ }

\newcommand{\ubvri}{\hbox{$U\!BV\!RI\!$} }
\newcommand{\bvri}{\hbox{$BV\!RI\!$} }
\newcommand{\nic}{\hbox{ $^{56}$Ni}}
\newcommand{\cob}{\hbox{ $^{56}$Co}}
\newcommand{\dmb}{\hbox{$\Delta m_{15}(B)$}}

\newcommand{\kms}{km~s$^{-1}$}
\newcommand{\md}{mag~d$^{-1}$}
\newcommand{\about}{$\sim\!\!$~}
\newcommand{\hal}{H$\alpha$}
\newcommand{\es}{SN~2002es}
\newcommand{\cx}{SN~2002cx}
\newcommand{\hk}{SN~2005hk}
\newcommand{\bg}{SN~1991bg}
\newcommand{\bh}{SN~1999bh}
\newcommand{\by}{SN~1999by}
\newcommand{\cf}{SN~2005cf}

\begin{document}

\title{The Low-Velocity, Rapidly Fading Type Ia Supernova 2002es}
\author{
Mohan Ganeshalingam\altaffilmark{1,2}, 
Weidong Li\altaffilmark{1},  
Alexei V. Filippenko\altaffilmark{1}, 
Jeffrey M. Silverman\altaffilmark{1},
Ryan Chornock\altaffilmark{3},
Ryan J. Foley\altaffilmark{3,4}, 
Thomas Matheson\altaffilmark{5},
Robert P. Kirshner\altaffilmark{3},
Peter Milne\altaffilmark{6},
Mike Calkins\altaffilmark{3}, and
Ken J. Shen\altaffilmark{1,7,8}
}
\altaffiltext{1}{
Department of Astronomy,
University of California,
Berkeley, CA 94720-3411, USA.
}
\altaffiltext{2}{
Electronic correspondence: mganesh@astro.berkeley.edu
}
\altaffiltext{3}{
Harvard-Smithsonian Center for Astrophysics, 
60 Garden Street, Cambridge, 
MA 02138, USA.}
\altaffiltext{4}{
Clay Fellow.
}
\altaffiltext{5}{
National Optical Astronomy Observatory,
950 North Cherry Avenue, 
Tucson, AZ 85719, USA.
}
\altaffiltext{6}{
Steward Observatory, University of Arizona, 
933 North Cherry Avenue, 
Tucson, AZ 85721, USA.
}
\altaffiltext{7}{
Lawrence Berkeley National Laboratory, 
Berkeley, CA 94720, USA.
}
\altaffiltext{8}{
Einstein Fellow.
}
\begin{abstract}
\es\ is a peculiar subluminous Type Ia supernova (SN Ia) with a combination of observed characteristics never before seen in a SN Ia. At maximum light, \es\ shares spectroscopic properties with the underluminous \bg\ subclass of SNe~Ia, but with substantially lower expansion velocities (\about 6000 \kms) more typical of the peculiar \cx\ subclass. One month after maximum light, spectra of \es\ show low velocities as seen in \cx. Photometrically, \es\ differs from both \bg-like and \cx-like supernovae. Although at maximum light it is subluminous  ($M_{B} = -17.78$ mag), \es\ has a relatively broad light curve ($\Delta m_{15}(B) = 1.28\pm0.04$ mag), making it a significant outlier in the light-curve width vs. luminosity relationship (the Phillips relation). From Arnett's law, we estimate a \nic\  mass of $0.17 \pm 0.05$  \Msun synthesized in the explosion, relatively low for a SN~Ia. One month after maximum light, we find an unexpected plummet in the bolometric luminosity. The late-time decay of the light curves is inconsistent with our estimated \nic\ mass, indicating that either the light curve was not completely powered by \nic\ decay or the ejecta became optically thin to $\gamma$-rays within a month after maximum light. The host galaxy is classified as an S0 galaxy with little to no star formation, indicating the progenitor of  \es\ is likely from an old stellar population. We also present a less extensive dataset for \bh, an object which shares similar photometric and spectroscopic properties. Both objects were found as part of the Lick Observatory Supernova Search, allowing us to estimate that these objects should account for 2.5\% of SNe~Ia within a fixed volume. We investigate theoretical models to probe the nature of \es, but find that current models are unable to explain all of its characteristics.
\end{abstract}
\shorttitle{The Peculiar Type Ia SN~2002es}
\shortauthors{Ganeshalingam et al.}

\section{Introduction}
Type Ia supernovae (SNe~Ia) are the runaway thermonuclear explosions of carbon-oxygen white dwarfs. SNe~Ia are characterized spectroscopically by an absence of hydrogen and the presence of intermediate-mass elements (e.g., silicon, sulfur, oxygen) and iron-group elements (iron, cobalt) \citep[e.g.,][and references therein]{filippenko97a}. A majority of spectroscopically identified SNe~Ia form a class of objects with a standardizable luminosity, allowing for their use as distance indicators. Application of SNe~Ia on extragalactic scales led to the discovery that the Universe is accelerating in its expansion \citep{riess98a,perlmutter99a}. Subsequent application of large samples of SNe~Ia out to high redshifts \citep{wood-vasey07a,hicken09b,kessler09b,amanullah10a,sullivan11b,suzuki12a} has led to precise estimates of cosmological parameters when combined with measurements of baryon acoustic oscillations (BAO) and anisotropy in the cosmic microwave background (CMB). 

The cosmological application of SNe~Ia is predicated on the relationship between the peak absolute magnitude of a SN, the width of its light curve, and its color. \cite{phillips93a} found that SNe with slowly declining light curves had a larger luminosity at maximum light. Applying corrections for light-curve width and SN color (as a measurement of host-galaxy extinction and intrinsic scatter in SN colors) has allowed SNe~Ia to be accurate distance indicators to within 10\% in distance \citep{jha07a,guy07a,conley08a}. There are indications that including spectral information \citep{foley08a,wang09b,bailey09a,foley11a,blondin11a,silverman12c} and host-galaxy information \citep{kelly10a,sullivan10a,lampeitl10a} further improves distance estimates.

Despite the ability to standardize the luminosity of SNe~Ia based on observed light-curve properties, a significant fraction of SN~Ia events have peculiar characteristics, including some overluminous and underluminous objects \citep[e.g.,][]{filippenko97a}. In particular, \citet{filippenko92a} and \citet{leibundgut93a} found that optical light curves of \bg\ evolved rapidly, and its peak luminosity was \about 2 mag fainter than that of normal objects. Moreover, the maximum-light spectrum of \bg\ showed strong \ion{Ti}{2} absorption, indicating a relatively cool photosphere, and the expansion velocity at maximum light as measured from the absorption minimum in the blueshifted \ion{Si}{2} $\lambda$6355 feature was \about10,000 \kms, slightly lower than expansion velocities measured for normal SNe~Ia (\about11,000 -- 13,000 \kms) \citep{filippenko92a}. Since the initial identification of \bg\ as a subclass of SNe~Ia, many members belonging to the subclass have been identified by the SN community. Studies of the host-galaxy morphology indicate that \bg-like objects are found preferentially in early-type galaxies \citep{howell01a,ganeshalingam10a, li11a}, leading to the suggestion that the progenitors of \bg-like SNe come from old stellar populations. There is currently debate about the cosmological utility of \bg-like objects as standardizable candles \citep{jha07a,guy07a}.

More recently, a range of properties for peculiar subluminous SNe~Ia have been discovered. \cite{foley10b} presented evidence that SN~2006bt spectroscopically resembled \bg, but photometrically resembled a normal SN~Ia. SN~2006bt was discovered  at a projected distance of 33.7~kpc from the nucleus of its early-type host galaxy. The Palomar Transient Factory \citep[PTF;][]{law09a} has published data on two peculiar subluminous objects, both found at large distances from the likely host galaxy. PTF~09dav was an abnormally subluminous SN Ia ($M_{B} = -15.44$~mag) with a narrow light curve found 41 kpc from its host galaxy \citep{sullivan11a}. \cite{maguire11a} presented data on the subluminous  PTF~10ops which shared many similarities with SN~2006bt. PTF~10ops had a broad light curve and was found at a projected distance of 148 kpc from the nominal host. All objects had spectral features that match those of \bg, although PTF~09dav had particularly slow expansion velocities of 6100 \kms.

Objects similar to \cx\ \citep{filippenko03a,li03a} form another subclass of subluminous peculiar SNe~Ia. These objects have maximum-light spectra similar to those of overluminous objects like SN~1991T, characterized by weak \ion{Si}{2} $\lambda$6355 features and dominated by \ion{Fe}{3} lines indicating a hot photosphere. However, the expansion velocities of these objects at maximum light are \about 6000 \kms, indicating an explosion with low kinetic energy per unit mass. There appears to be a great diversity among \cx-like objects, with a distribution of absolute luminosity and kinetic energy \citep{narayan11a,mcclelland10a}. SN~2008ha is the faintest member of the subclass, with $M_{V} = -14.2$ mag and velocities of \about 4000--5000 \kms\ at maximum light \citep{foley10c}. 

Here we report our observations of \es, an object somewhat spectroscopically similar to \bg, further adding to the puzzle of subluminous SNe~Ia. \es\ was discovered \citep{li02a} on unfiltered CCD images at \about 16.3 mag as part of the Lick Observatory Supernova Search (LOSS)  with the 0.76-m Katzman Automatic Imaging Telescope \citep[KAIT;][]{filippenko01a,li03b} on 2002 Aug. 23.5 (UT dates are used throughout this paper) in UGC 2708. Its J2000 coordinates are $\alpha =03^{\rm h}23^{\rm m}47^{\rm s}\!.23$ and $\delta =+40^{\circ}33^{\rm m}53.5^{\rm s}$, which is 19$''$ W and 26$''$ N of the galaxy nucleus \citep{li02a}. Subsequent optical spectroscopic observations on 2002 Sep. 03  by \cite{chornock02a} and \cite{matheson02a} classified \es\ as a SN~1991bg-like object \citep{filippenko92a} near maximum light based on the presence of strong \ion{Ti}{2}, \ion{Si}{2}, and \ion{O}{2} absorption features, but with an expansion velocity of  \about 6000~\kms\ as measured from the absorption minimum of the \ion{Si}{2} $\lambda$6355 feature. This is notably lower than typical  expansion velocities of \about11,000~\kms\ found in normal SNe~Ia  and \bg-like objects and more typical of \cx-like objects \citep{li03a}.

Noting the peculiarity of \es, our group started a photometric and spectroscopic campaign to document the evolution of this unique object. We present \bvri\ photometry for \es\ starting a week before maximum light collected as part of the LOSS SN~Ia photometry program \citep{ganeshalingam10a}. Late-time photometry was obtained with the 2.3-m Bok telescope at Steward Observatory on Kitt Peak in Arizona to constrain the late-time decay of the light curve. We also present an extensive optical spectral series covering the evolution of \es\ from maximum light to two months after maximum. In addition, we present evidence that SN~1999bh is a \es-like event.

The rest of the paper is structured as follows. Our data and reduction techniques are presented in \S \ref{s:obs}, with a detailed analysis of our photometry and spectroscopy in \S \ref{s:results}. In \S \ref{s:discussion}, we discuss possible physical interpretations for the observed properties of \es\ and compare \es\ to theoretical models. We summarize our results  in \S \ref{s:conclusions}.

\begin{figure}[!t]
\begin{center}
\includegraphics[scale=.5]{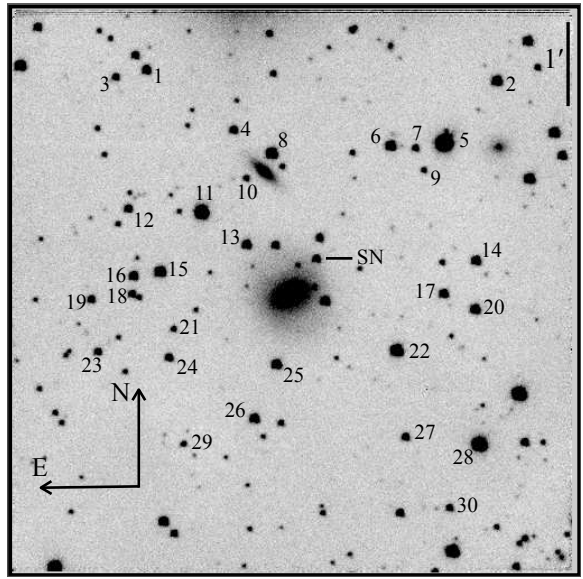}
\end{center}
{\caption{KAIT $V$-band image of  \es. The field is $6.7\arcmin \times 6.7\arcmin$. The SN and comparison stars are marked. The labels of the comparison stars correspond to the numbers in Table \ref{t:comp_star}.} \label{f:finder}}
\end{figure}

\section{Observations}
\label{s:obs}
\subsection{Photometry}
Broadband photometric follow-up observations of \es\ started on 2002 Aug. 24.48 (one week before $B_{\rm max}$) with KAIT in the \bvri\ bands. \es\ was monitored with a 1--2 day cadence for the first month after discovery, resulting in well-sampled light curves. Additional \bvri\ photometry from the 1-m Nickel Telescope at Lick Observatory was obtained to complement the KAIT data. Late-time data in $BV\!R$ were taken using the 2.3-m Bok telescope. A $V$-band image  of the field from KAIT is shown in Figure \ref{f:finder}.

Images were bias subtracted and flat fielded using standard procedures in \verb IRAF .\footnote{IRAF: The Image Reduction and Analysis Facility is distributed by the National Optical Astronomy Observatory, which is operated by the Association of Universities for Research in Astronomy (AURA), Inc., under cooperative agreement with the National Science Foundation (NSF).} \es\ was sufficiently close to its host galaxy that galaxy subtraction was required to disentangle SN light from galaxy light. Templates for the field were obtained for the KAIT, Nickel, and Bok telescopes after \es\ faded beyond the detection limit of the telescope ($ >$2 yr after discovery) and subtracted from the data images. The flux from the SN was measured in comparison to local field stars using point-spread function (PSF) fitting photometry. The uncertainties in our photometric measurements were estimated by randomly injecting artificial stars with the same PSF and magnitude as the SN into the data images. The images with artificial stars were rerun through our galaxy-subtraction and photometry routines, and the scatter in doing this 20 times was adopted as the photometric measurement error.

Instrumental magnitudes were color-corrected to the Landolt system \citep{landolt92a,landolt09a} using the average color terms measured from multiple photometric nights. The magnitudes of local field stars were calibrated against Landolt standards on six photometric nights. The results are reported in Table \ref{t:comp_star}. The final photometric error is the photometric measurement error added in quadrature with the error in the calibration of our field stars. The final photometry is presented in Table \ref{t:phot}.

\begin{figure}[!t]
\begin{center}
\includegraphics[scale=.5]{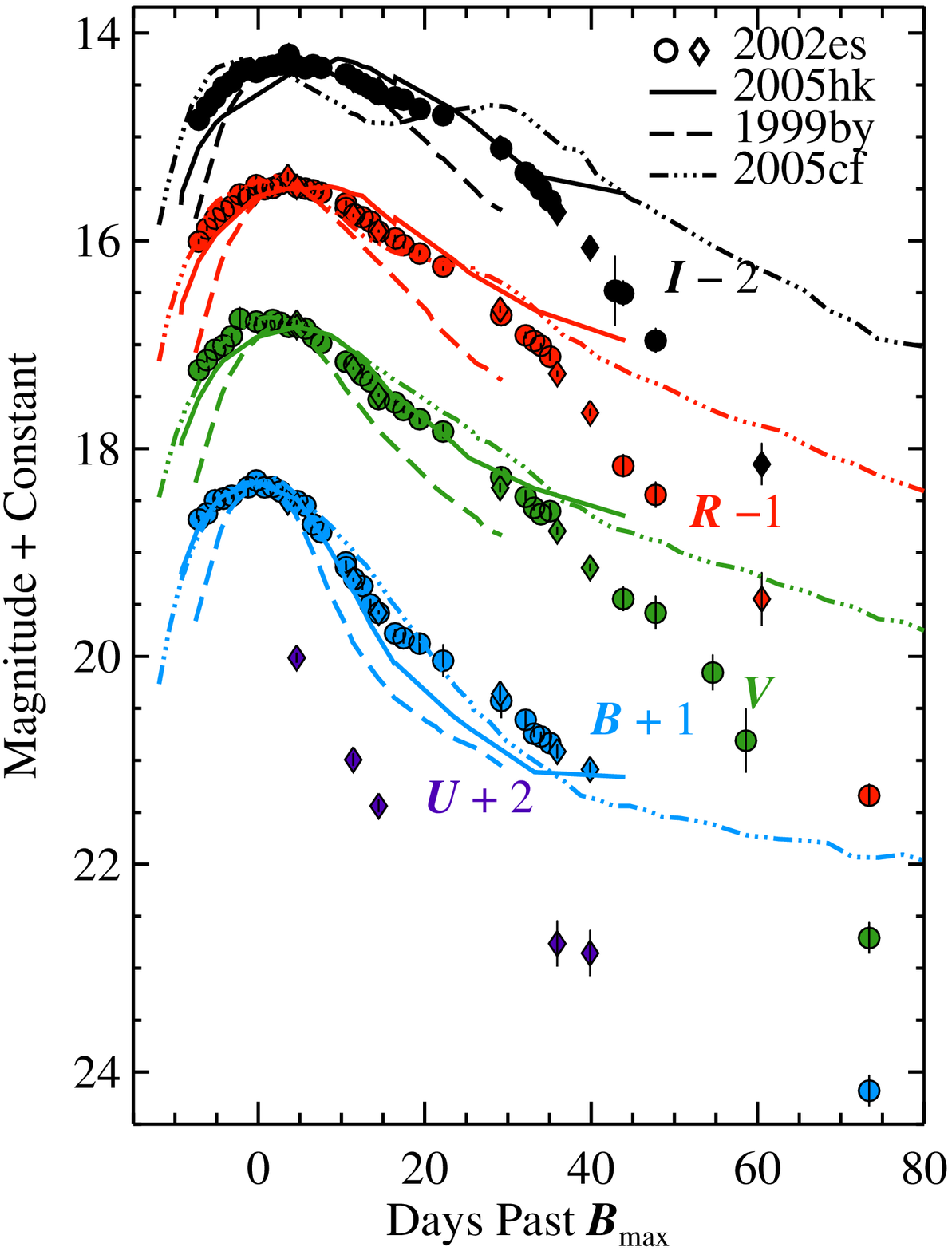}
\end{center}
{\caption{\ubvri light curves of SN 2002es. Data from the KAIT, Nickel, and Bok telescopes are plotted as circles. Data from \cite{hicken09a} are plotted as diamonds.  For comparison, we plot the \cx-like  \hk\ (solid line), the \bg-like \by\ (dashed), and the normal \cf\ (dot-dashed line). Comparison light curves have been shifted to have the same peak magnitude and phase as \es. Note the particularly fast decline in all bands at $t > +30$~d compared to the other objects.} \label{f:lc}}
\end{figure}

In addition to the photometry presented here, we supplement our light curves with \ubvri\!\! data presented for \es\ in the CfA3 photometry sample \citep{hicken09a} taken with the 1.2-m telescope at Fred Lawrence Whipple Observatory (FLWO) in Arizona operated by the Harvard Smithsonian Center for Astrophysics (CfA). The agreement between the KAIT and CfA3 data is within $\sim 0.05$~mag in all bands, with no evidence of a systematic offset. The light curves from all telescopes are presented in Figure \ref{f:lc}.

\subsection{Spectroscopy {\label{ss:spec}}}
Low-resolution spectra of \es\ were obtained using the Kast dual spectrograph mounted on the 3-m Shane Telescope at Lick observatory \citep{miller93a}, the FAST spectrograph mounted on the 1.5-m Tillinghast telescope at FLWO \citep{fabricant98a}, and LRIS on the 10-m Keck I telescope \citep{oke95a}. All of our observations were taken at the optimal parallactic angle to minimize differential light loss \citep{filippenko82a}.

All spectra of \es\ were reduced using standard CCD processing techniques \citep[e.g.,][]{foley03a,matheson08a}. Processing and extraction of the one-dimensional spectrum were performed in \verb IRAF  using the optimal extraction algorithm of \cite{horne86a}. The wavelength calibration was obtained by fitting low-order polynomials to calibration-lamp spectra. Our spectra were flux calibrated using our own \verb IDL  routines.  Corrections for telluric absorption features were made using spectrophotometric standards \citep{wade88a} taken at roughly the same airmass as the SN observation. Table \ref{t:spec} presents a summary of our spectroscopic observations.

\begin{deluxetable*} {c c c c c c c c}
\tablewidth{0pc}
\tablecaption{Photometry of Local Standard Stars \label{t:comp_star}}
\tablehead { \colhead{ID} & 
             \colhead{$\alpha$ (J2000)}  &
             \colhead{$\delta$ (J2000)}  &
             \colhead{$B (\sigma_{B})$ (mag)}   &
             \colhead{$V (\sigma_{V})$ (mag)}   &
             \colhead{$R (\sigma_{R})$ (mag)}   &
             \colhead{$I (\sigma_{I})$ (mag)}   &
             \colhead{$N_{\rm calib}$}}
\startdata
 1  &  03:23:57.81  &  +40:36:08.0  &  17.403 (011)  &  16.497 (008)  &  15.940 (012)  &  15.462 (007)  &  4 \\
 2  &  03:23:35.94  &  +40:36:00.3  &  16.470 (010)  &  15.675 (007)  &  15.206 (007)  &  14.744 (012)  &  5 \\
 3  &  03:23:59.75  &  +40:36:02.8  &  17.817 (010)  &  17.174 (011)  &  16.785 (012)  &  16.331 (014)  &  3 \\
 4  &  03:23:52.38  &  +40:35:25.2  &  17.440 (010)  &  16.844 (007)  &  16.482 (011)  &  16.104 (011)  &  4 \\
 5  &  03:23:39.24  &  +40:35:16.1  &  14.293 (009)  &  12.933 (014)  &  12.191 (002)  &  11.545 (011)  &  2 \\
 6  &  03:23:42.58  &  +40:35:14.0  &  16.442 (007)  &  15.702 (003)  &  15.265 (005)  &  14.851 (011)  &  5 \\
 7  &  03:23:41.02  &  +40:35:12.2  &  17.978 (011)  &  17.314 (011)  &  16.905 (013)  &  16.482 (010)  &  5 \\
 8  &  03:23:50.00  &  +40:35:08.6  &  15.777 (011)  &  15.119 (005)  &  14.727 (007)  &  14.325 (011)  &  4 \\
 9  &  03:23:40.51  &  +40:34:56.4  &  18.618 (013)  &  17.970 (004)  &  17.579 (010)  &  17.035 (008)  &  2 \\
10  &  03:23:51.59  &  +40:34:50.8  &  18.657 (013)  &  17.786 (006)  &  17.292 (008)  &  16.865 (006)  &  3 \\
11  &  03:23:54.37  &  +40:34:26.8  &  14.523 (012)  &  14.040 (011)  &  13.736 (013)  &  13.407 (006)  &  4 \\
12  &  03:23:58.96  &  +40:34:29.2  &  17.496 (012)  &  16.721 (011)  &  16.254 (013)  &  15.735 (013)  &  4 \\
13  &  03:23:51.57  &  +40:34:03.6  &  16.938 (010)  &  16.180 (005)  &  15.744 (011)  &  15.296 (015)  &  4 \\
14  &  03:23:37.25  &  +40:33:52.3  &  17.115 (009)  &  16.126 (006)  &  15.542 (007)  &  15.047 (012)  &  4 \\
15  &  03:23:56.96  &  +40:33:44.5  &  15.783 (012)  &  15.258 (012)  &  14.905 (007)  &  14.554 (010)  &  4 \\
16  &  03:23:58.62  &  +40:33:41.4  &  16.993 (010)  &  16.280 (011)  &  15.833 (008)  &  15.395 (015)  &  4 \\
17  &  03:23:39.26  &  +40:33:28.8  &  16.795 (010)  &  16.222 (005)  &  15.863 (008)  &  15.503 (012)  &  4 \\
18  &  03:23:58.71  &  +40:33:28.4  &  17.935 (011)  &  17.021 (003)  &  16.477 (008)  &  16.003 (007)  &  4 \\
19  &  03:24:01.27  &  +40:33:24.7  &  18.038 (013)  &  16.979 (010)  &  16.402 (008)  &  15.814 (017)  &  3 \\
20  &  03:23:37.30  &  +40:33:17.8  &  16.652 (013)  &  15.842 (011)  &  15.360 (009)  &  14.935 (013)  &  4 \\
21  &  03:23:56.13  &  +40:33:03.5  &  18.860 (010)  &  17.906 (005)  &  17.374 (012)  &  16.956 (008)  &  3 \\
22  &  03:23:42.20  &  +40:32:48.4  &  15.821 (011)  &  14.705 (003)  &  14.090 (005)  &  13.512 (009)  &  5 \\
23  &  03:24:00.85  &  +40:32:47.5  &  17.769 (009)  &  17.064 (005)  &  16.658 (006)  &  16.233 (005)  &  2 \\
24  &  03:23:56.43  &  +40:32:43.2  &  17.724 (011)  &  16.895 (004)  &  16.377 (009)  &  15.836 (013)  &  5 \\
25  &  03:23:49.70  &  +40:32:38.5  &  16.919 (009)  &  15.795 (004)  &  15.101 (007)  &  14.540 (003)  &  4 \\
26  &  03:23:51.07  &  +40:31:60.0  &  16.680 (011)  &  15.987 (005)  &  15.583 (007)  &  15.170 (013)  &  5 \\
27  &  03:23:41.64  &  +40:31:46.7  &  18.034 (012)  &  16.877 (011)  &  16.157 (012)  &  15.481 (014)  &  2 \\
28  &  03:23:37.03  &  +40:31:41.7  &  14.570 (007)  &  13.860 (007)  &  13.427 (012)  &  13.003 (016)  &  3 \\
29  &  03:23:55.50  &  +40:31:41.6  &  18.669 (011)  &  17.703 (012)  &  17.062 (011)  &  16.456 (007)  &  4 \\
30  &  03:23:38.90  &  +40:30:56.3  &  17.954 (013)  &  17.261 (010)  &  16.836 (013)  &  16.417 (014)  &  4 
\enddata
\tablecomments{1$\sigma$ uncertainties (in units of 0.001 mag) are listed in parentheses.}
\end{deluxetable*}

 \begin{deluxetable*}{cccccc}
 \tablewidth{0pc}
 \tablecaption{Photometry of SN 2002es \label{t:phot}}
 \tablehead{\colhead{JD $-$ 2,452,000} & \colhead{$B$ (mag)} & \colhead{$V$ (mag)}& \colhead{$R$ (mag)}& \colhead{$I$ (mag)} & \colhead{Telescope}}
 \startdata
    510.98 & 17.682  (033) & 17.245  (028) & 17.007  (025) & 16.835  (031) & KAIT \\
    511.98 & 17.625  (029) & 17.148  (027) & 16.882  (022) & 16.711  (029) & KAIT \\
    512.97 & 17.498  (033) & 17.050  (028) & 16.786  (027) & 16.622  (030) & KAIT \\
    514.00 & 17.482  (034) & 17.009  (029) & 16.711  (029) & 16.521  (033) & KAIT \\
    515.01 & 17.453  (035) & 16.918  (026) & 16.669  (020) & 16.470  (030) & KAIT \\
    516.01 & $\cdots$      & 16.752  (111) & 16.552  (097) & 16.375  (033) & KAIT \\
    517.02 & 17.372  (104) & $\cdots$      & 16.587  (055) & 16.347  (046) & KAIT \\
    518.01 & 17.307  (044) & 16.778  (045) & 16.465  (090) & 16.374  (022) & KAIT \\
    519.02 & 17.373  (020) & 16.817  (024) & 16.503  (020) & 16.332  (026) & KAIT \\
    520.01 & 17.369  (020) & 16.762  (020) & 16.493  (022) & 16.318  (026) & KAIT \\
    520.99 & 17.414  (026) & 16.790  (020) & 16.453  (020) & 16.303  (026) & KAIT \\
    521.99 & $\cdots$      & 16.829  (031) & $\cdots$      & 16.211  (111) & KAIT \\
    523.02 & 17.506  (023) & 16.829  (021) & 16.484  (020) & 16.320  (031) & KAIT \\
    524.02 & 17.550  (026) & 16.844  (020) & 16.498  (020) & 16.339  (037) & KAIT \\
    524.95 & 17.729  (020) & 16.925  (020) & 16.514  (020) & 16.311  (020) & 1~m\\
    525.95 & 17.805  (020) & 16.985  (020) & 16.539  (020) & 16.336  (020) & 1~m\\
    528.97 & 18.094  (020) & 17.165  (020) & 16.687  (020) & 16.400  (020) & 1~m\\
    529.00 & 18.142  (030) & 17.164  (028) & 16.640  (023) & 16.400  (023) & KAIT \\
    530.01 & 18.255  (040) & 17.193  (025) & 16.746  (020) & 16.446  (023) & KAIT \\
    531.01 & 18.323  (039) & 17.293  (031) & 16.774  (020) & 16.492  (022) & KAIT \\
    531.99 & 18.500  (049) & 17.354  (020) & 16.814  (020) & 16.533  (029) & KAIT \\
    533.00 & 18.583  (071) & 17.525  (080) & 16.914  (105) & 16.588  (101) & KAIT \\
    535.00 & 18.779  (037) & 17.556  (031) & 16.977  (033) & 16.621  (025) & KAIT \\
    536.02 & 18.824  (053) & 17.629  (032) & 17.042  (033) & 16.641  (028) & KAIT \\
    537.99 & 18.876  (094) & 17.717  (050) & 17.122  (030) & 16.732  (043) & KAIT \\
    540.87 & 19.041  (155) & 17.836  (063) & 17.248  (043) & 16.792  (058) & KAIT \\
    547.98 & 19.429  (167) & 18.275  (084) & 17.715  (036) & 17.110  (055) & KAIT \\
    550.97 & 19.610  (082) & 18.464  (061) & 17.910  (032) & 17.346  (033) & KAIT \\
    551.99 & 19.744  (032) & 18.570  (025) & 17.964  (020) & 17.417  (042) & 1~m\\
    552.84 & 19.771  (043) & 18.631  (027) & 18.010  (021) & 17.505  (025) & 1~m\\
    553.96 & 19.834  (100) & 18.602  (070) & 18.116  (038) & 17.615  (058) & KAIT \\
    561.91 & $\cdots$      & $\cdots$      & $\cdots$      & 18.481  (334) & KAIT \\
    562.91 & $\cdots$      & 19.447  (118) & 19.167  (109) & 18.508  (123) & KAIT \\
    566.87 & $\cdots$      & 19.579  (161) & 19.445  (125) & 18.961  (120) & KAIT \\
    573.86 & $\cdots$      & 20.155  (171) & $\cdots$      & $\cdots$      & KAIT \\
    577.92 & $\cdots$      & 20.811  (307) & $\cdots$      & $\cdots$      & KAIT \\
   592.89 &  23.180  (150) & 22.711 (110)       & 22.34 (110) & $\cdots$      & Bok 
 \enddata
\tablecomments{1$\sigma$ uncertainties (in units of 0.001 mag) are listed in parentheses.}
 \end{deluxetable*}

\subsection{Host Galaxy \label{ss:host}} 
The NASA/IPAC Extragalactic Database (NED)\footnote{http://ned.ipac.caltech.edu/ .} lists two discordant values for the heliocentric redshift of UGC 2708 from different sources. The NED webpage for UGC 2708 lists  $z_{\rm helio} = 0.028$ as the default redshift determined from marginal measurements of the \ion{H}{1} 21-cm line by \cite{monnier-ragaigne03a}. NED warns under the ``Essential Note"  section that \cite{de-vaucouleurs91a} find $z_{\rm helio} = 0.018$ from optical lines. We obtained a spectrum of UGC 2708 with Kast at Lick Observatory on 2006 July 21.4 to determine the actual redshift of the host galaxy. We find a heliocentric redshift of $z_{\rm hel} = 0.0182 \pm 0.0001$ from measurements of weak, narrow H$\alpha$ + [\ion{N}{2}]  emission lines, in good agreement with the value of \cite{de-vaucouleurs91a}. The heliocentric redshift of UGC 2708 corresponds to $z_{\rm CMB} = 0.0177$ (in the frame of the cosmic microwave background). We adopt an uncertainty of 300 \kms\ to account for any peculiar motions induced by gravitational interactions with neighboring galaxies. For a standard $\Lambda$CDM cosmology with $\Omega_{\rm m} = 0.27$, $\Omega_{\Lambda}$ = 0.73, $w = -1$, and $H_{0} = 73.8~{\rm km~s^{-1}~Mpc^{-1}}$ \citep{riess11b}, we find a luminosity distance $d_{L} = 73.20 \pm 4.41$ Mpc and a distance modulus $\mu = 34.32 \pm 0.12$ mag.  \es\ exploded at a projected distance of 11 kpc  from the nucleus of UGC 2708.

The original IAUCs that spectroscopically classify \es\ \citep{chornock02a,matheson02a} predate the source for the incorrect redshift listed on NED and likely used the correct redshift supplied by \cite{de-vaucouleurs91a}. The correct redshift for \es\ was also used in the cosmology analysis of \cite{hicken09b}.

UGC 2708 was included in a study of SN~Ia host-galaxy properties by \cite{neill09a}. The authors estimated galaxy properties by fitting template-galaxy spectral energy distributions (SEDs) to multi-wavelength photometry from the Sloan Digital Sky Survey (SDSS) and the Galaxy Evolution Explorer (GALEX). For UGC 2708, the authors estimated a minimal amount of active star formation and $E(B-V)_{\rm host} = 0.0$~mag.  This finding, along with the absence of \ion{Na}{1} D absorption at the redshift of the host galaxy in spectra of \es, indicate that the reddening due to the host galaxy is negligible.

UGC 2708 was also observed spectroscopically as part of SDSS. Using the publicly available line-flux measurements, we find an [\ion{N}{2}]/H$\alpha$ ratio consistent with that of a low-ionization nuclear emission-line region \citep[LINER;][]{heckman80a} or a composite galaxy. Coupled with the results from \cite{neill09a}, UGC 2708 is likely a LINER with no active star formation.

 \begin{deluxetable*}{ccccc}
 \tablewidth{0pc}
 \tablecaption{Log of Optical Spectral Observations for \es \label{t:spec}}
 \tablehead{\colhead{UT Date} & \colhead{Phase\tablenotemark{a} (d)} & \colhead{Telescope/Instrument}& \colhead{Exposure Time (s)}& \colhead{Observer\tablenotemark{b}}}
 \startdata
2002 Sep. 03.5   & +3  & Lick/Kast & 600 & AF, RC, BS \\
2002 Sep. 05.5   & +5 & FLWO/FAST & 1200 & MC \\
2002 Sep. 06.5   & +6  & FLWO/FAST & 1200 & MC \\
2002 Sep. 10.5   & +10 & FLWO/FAST & 1200 & PB \\
2002 Sep. 12.5   & +12 & FLWO/FAST & 1200 & MC \\
2002 Sep. 13.3   & +13 & Lick/Kast & 1200  & RF, SJ, MP \\
2002 Sep. 28.4   & +28 & FLWO/FAST & 1200 & MC \\
2002 Oct. 01.3   & +30 & Lick/Kast &1800 & AF,RF \\
2002 Oct. 08   & +37 & Keck I/LRIS  & 400 & AF,RC \\
2002 Nov. 08   & +67 & Keck I/LRIS  & 1800 & AF, RC,SJ, BB \\
2002 Nov. 11   & +70 & Keck I/LRIS & 1800 & AF, RC 
\enddata
\tablenotetext{a}{Rest-frame days relative to the date of $B_{\rm max}$, 2002 Aug. 31.8 (JD 2,452,518.3), rounded to the nearest day.}
\tablenotetext{b}{AF = A. Filippenko, BB = B. Barris, BS = B. Swift, MC = M. Calkins, MP = M. Papenkova, PB = P. Berlind,  RC = R. Chornock, RF = R. Foley, SJ = S. Jha}.
\end{deluxetable*}

\section{Results }
\label{s:results}
\subsection{Photometry \label{s:phot}}
The light curves of \es\ are displayed in Figure \ref{f:lc}, along with light curves of the normal SN~Ia 2005cf \citep{wang08a,ganeshalingam10a}, the \cx-like SN~Ia 2005hk \citep{phillips07a}, and the \bg-like SN~Ia 1999by \citep{ganeshalingam10a}.

Basic photometric properties for the light curves of \es\  are reported in Table \ref{t:phot_prop}. All values were measured using a fifth-order polynomial fit directly to the data. Uncertainties were estimated using a Monte Carlo routine to produce 50 realizations of our dataset. Each individual dataset realization was produced by randomly perturbing each photometry data point using its photometric uncertainty assuming a Gaussian distribution. Light-curve properties were measured for each realization, and the final measurements were found by taking the mean and standard deviation of the set of simulated realizations. 

The light curves of \es\ share some characteristics in common with the subluminous \bg\ subtype. \es\ lacks a prominent shoulder in the $R$ and $I$ bands \citep{filippenko92a,leibundgut93a}. The secondary maximum often found in $R$ and $I$ (and more prominently in the near-infrared) is attributed to the cooling of the ejecta to temperatures where the transition from \ion{Fe}{3} to  \ion{Fe}{2} becomes favorable, redistributing flux from shorter wavelengths to longer wavelengths \citep{kasen06a}. This transition occurs earlier in cooler SNe. From models of the radiative transfer within SNe, \cite{kasen06a} finds that the timing and strength of the shoulder is dependent on the distribution and amount of \nic\ within the ejecta. Models with a completely homogenized composition and with a small amount of \nic\ result in an $I$-band light curve with no discernible secondary peak or shoulder. Instead, the two peaks merge to produce a single broad peak. Given the similar absolute magnitudes of \es\ and \bg, the lack of a shoulder or secondary maximum in the $R$ and $I$ bands may be a consequence of low \nic\ production, if the light curve is powered by the decay of \nic.

The timing of maximum light in each band is similar to that of \bg-like SNe. In normal SNe, maximum light in the $I$ band precedes that of $B$ by a few days (as evident in the light curves of SN~2005cf in Figure \ref{f:lc}). \cite{taubenberger08a} find that for the \bg-like SN~2005bl, peak brightness in \ubvri\ occurred in successive order of bluest to reddest filter with the date of maximum light in each band separated by \about 1~d, similar to what is seen in \es.

The $B$-band peak magnitude of  \es\  is $17.33 \pm 0.02$~mag after correcting for a color excess of $E(B-V) = 0.183$~mag from Milky Way extinction \citep{schlegel98a}. We assume negligible host-galaxy extinction based on the classification of UGC 2708 as an S0 galaxy and no evidence of \ion{Na}{1} D absorption at the host-galaxy redshift. This corresponds to an absolute magnitude of $M_{B} = -17.78 \pm 0.12$~mag,  comparable to other subluminous SNe Ia, but brighter than \bg-like objects \citep{taubenberger08a}. The absolute magnitude of \es\ is brighter than that of the \bg-like SN~2005bl by \about 0.5--1 mag in \bvri.

\begin{deluxetable*}{ccccc}
\tablewidth{0pc}
\tablecaption{Photometric Properties of \es \label{t:phot_prop}}
\tablehead{\colhead{Filter} & \colhead{JD of max $-$ 2,452,000} & \colhead{Mag at max\tablenotemark{a}} & \colhead{Peak abs. mag\tablenotemark{b}}  & \colhead{$\Delta m_{15}$ (mag)}}
\startdata
$B$  &  $518.30 \pm 0.28$ & $17.33 \pm 0.02 $  & $-17.78 \pm 0.12 $  & $1.28 \pm 0.04 $  \\
$V$  &  $519.21\pm  0.28$  & $ 16.78 \pm 0.02$  & $-18.15 \pm 0.12$  &  $0.74 \pm 0.02 $ \\
$R$ &  $520.96 \pm 0.30$  & $16.47 \pm 0.02$   & $-18.35 \pm 0.12$  &   $0.57 \pm 0.02$ \\
$I $  & $521.39\pm 0.33$   & $16.29 \pm 0.02$   &  $-18.40 \pm 0.12$ &   $0.37 \pm 0.02$ 
\enddata
\tablenotetext{a}{Not corrected for Milky Way or host-galaxy extinction.}
\tablenotetext{b}{Corrected for Milky Way extinction and assuming no host-galaxy extinction.}
\end{deluxetable*}

Despite these similarities with \bg-like SNe~Ia, the light curves of \es\ are significantly broader in all bands compared to those of most subluminous objects. Using the decline in magnitudes between maximum light and 15 days after maximum light in the $B$ band as a proxy for light-curve width \citep{phillips93a}, we measure $\Delta m_{15} (B) = 1.28 \pm 0.04~\rm{mag}$. Other \bg-like SNe typically have \dmb\ $\approx 1.9$~mag \citep{taubenberger08a}.

The light curves of \es\ also share similarities to the \cx-like \hk. Both objects are subluminous events compared to normal SNe~Ia with light curves that are broader than those of \bg-like objects and lack a shoulder in the $R$ and $I$ bands. \es\  has a slower $B$-band decline than most \cx-like objects. \hk\ had $\Delta m_{15}(B) = 1.56 \pm 0.09$ mag.

As the outer ejecta turn transparent, the light curve is expected to be powered by the thermalization of $\gamma$-rays produced by the decay of $^{56}\rm{Co}$ (\about 50~d after explosion). This is typically observed as a linear fading in all bands. \cite{leibundgut00a} find typical decay rates for SNe~Ia of 0.014 \md\ in $B$, 0.028  \md\ in $V$, and 0.042  \md\ in $I$. For \bg, the $B$-band decline is marginally faster at 0.019 \md\ and slower in $I$ at 0.040 \md. The $V$-band decline has been found to be fairly constant among normal and \bg-like SNe~Ia \citep{leibundgut00a}.

Using data at $t > +30$~d (relative to maximum light in $B$ band), we measure a decline of $0.040 \pm 0.004$ \md\ in $B$, $0.081 \pm 0.004$ \md\ in $V$, $0.101 \pm 0.004$ \md\ in $R$, and $0.099 \pm 0.004$ \md\ in $I$. These rates are substantially faster than expected for an object powered by the decay of $^{56}$Co, even after accounting for the declining $\gamma$-ray deposition function as the ejecta expand homologously, and cast doubt on whether \es\ is necessarily a thermonuclear explosion. In \S \ref{ss:late}, we discuss possible explanations for such a fast decline.

\begin{figure}[!t]
\begin{center}
\includegraphics[scale=.43]{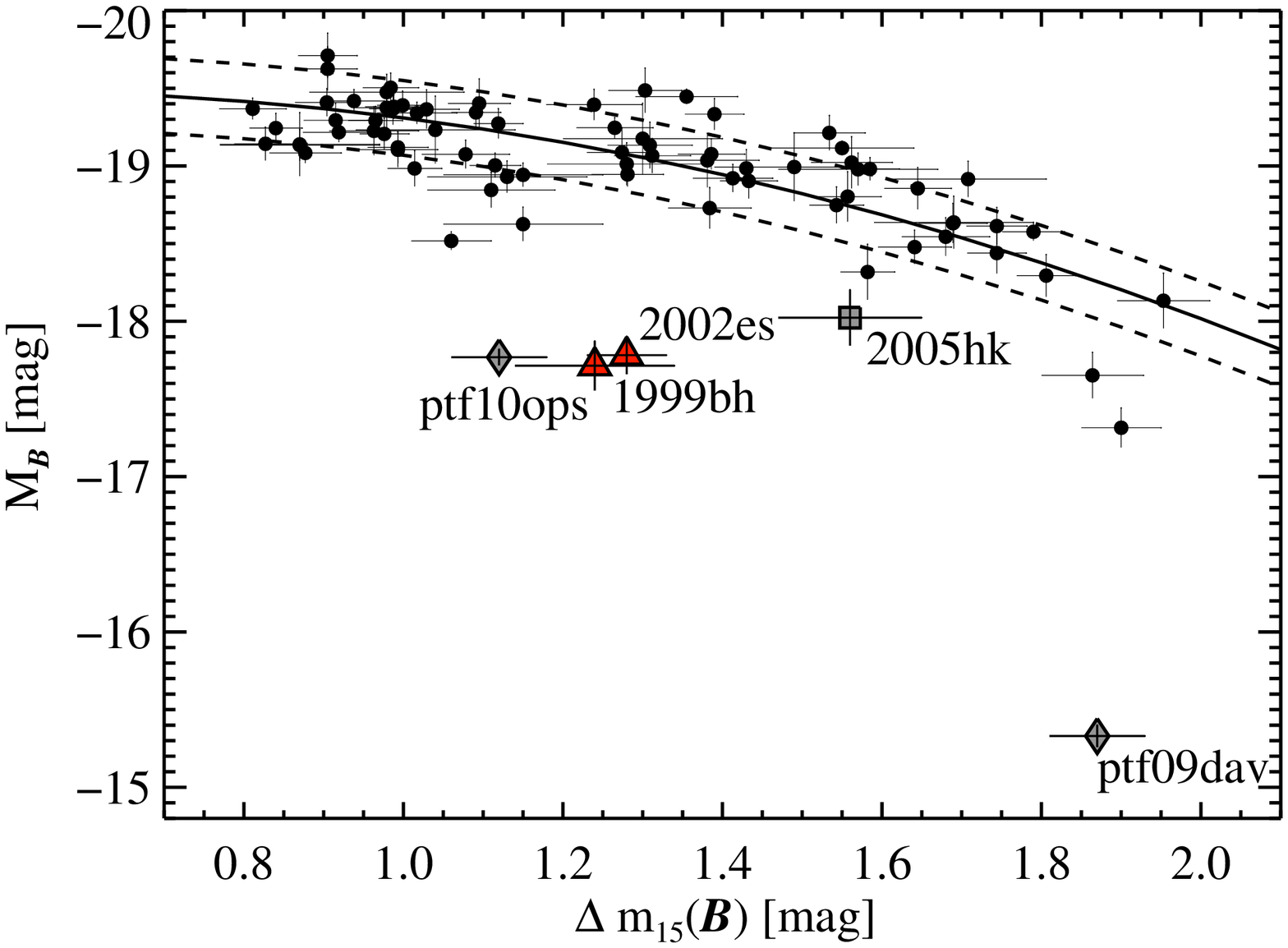}
\end{center}
{\caption{$M_{B}$ as a function of \dmb\ for 76 SNe from \cite{ganeshalingam10a} corrected for host-galxay extinction using values derived from the MLCS distance-fitting software. Overplotted as a solid line is the quadratic Phillips relation from \cite{phillips99a} shifted along the ordinate to match the data. The dashed lines correspond to the $1\sigma$ scatter about the relationship. \es\ and \bh\ (red triangles) are clear outliers in the relationship. For comparison, we include other subluminous peculiar objects \hk, PTF~09dav, and PTF~10ops.} \label{f:phillips}}
\end{figure}

In Figure \ref{f:phillips}, we plot $\Delta m_{15}(B)$ vs. $M_{B}$ for \es\ along with 76 SNe taken from \cite{ganeshalingam10a} using host-galaxy extinction values determined using the Multicolor Light-Curve Shape method \citep[MLCS2k2.v006;][]{jha07a}. The light-curve width vs. luminosity relationship from \cite{phillips99a}, adjusted to $H_{0} = 73.8~{\rm km~s^{-1}~Mpc^{-1}}$, is overplotted as a solid line with $1\sigma$ scatter about the relation indicated by dashed lines. We also include measurements for the peculiar SNe SN~2005hk \citep{phillips07a}, PTF~09dav \citep{sullivan11a}, and PTF~10ops \citep{maguire11a}. \es\ is an obvious outlier ($5 \sigma$) in the Phillips relation which predicts that \es\ should be \about 1.3 mag brighter than the observed peak magnitude. The amount of host-galaxy extinction required to explain this discrepancy is unlikely given the absence of \ion{Na}{1} D absorption in \es\ spectra at the redshift of the host galaxy. 

\subsection{Color Curves}
\begin{figure}[!t]
\begin{center}
\includegraphics[scale=.55]{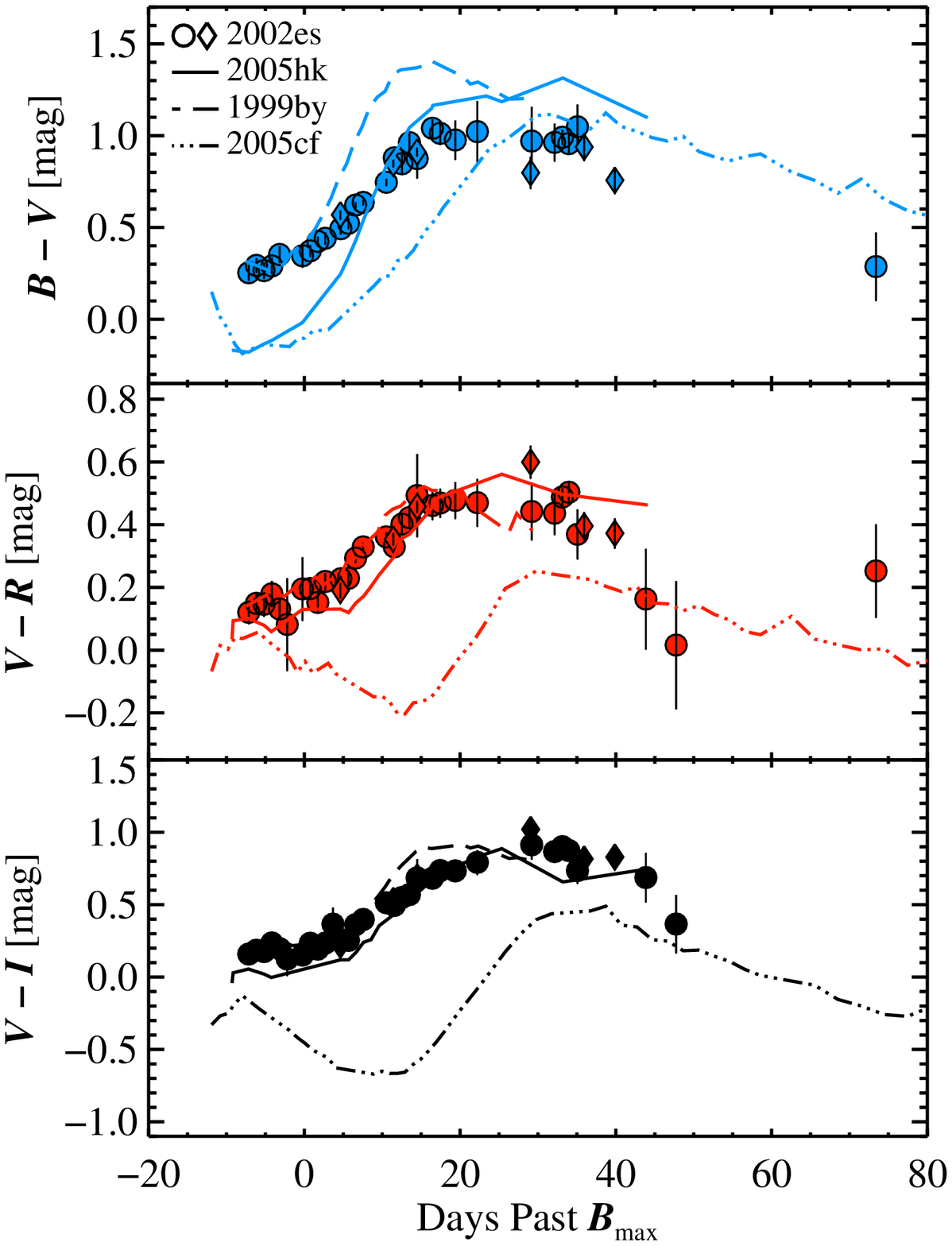}
\end{center}
{\caption{Optical color curves of \es\ corrected for $E(B-V) = 0.183$~mag from Milky Way extinction. Solid circles are data taken with the KAIT and Nickel telescopes and solid diamonds are from \cite{hicken09a}. Shown for comparison are \hk, \by, and \cf\ corrected for extinction using the reddening values provided in their respective reference. \es\ is significantly redder in all colors compared to \cf. None of the comparison color curves provides an adequate match to \es.} \label{f:col}}
\end{figure}

In Figure \ref{f:col}, we plot the color evolution of \es\ along with \cf, \hk, and \by\ for comparison. All objects have been corrected for Milky Way extinction using the dust maps of \cite{schlegel98a}. We adopt a host-galaxy color excess of $E(B-V)_{\rm host} = 0.10$~mag for \cf\ \citep{wang09a}, $E(B-V)_{\rm host} = 0.10$~mag for \hk\ \citep{chornock06a}, and $E(B-V)_{\rm host} = 0.0$~mag for \by\ \citep{garnavich04a}. We assume no host-galaxy extinction for \es.

The $B-V$ color curve of \es\ does not match that of any of our comparison objects. The color evolution is most similar to that of the \bg-like \by, but with significant differences in $B-V$ after maximum light in $B$. Before $t(B_{\rm max})$, \es\ and \by\ share a similar $B-V$ color evolution that is considerably redder than that of \cf\ and \hk. However, at \about 5 d past maximum light in $B$, \by\ quickly becomes redder in $B-V$ while the color evolution of \es\ is much more gradual. The $V-R$ and $V-I$ color evolution for \es\ and \by\ appear much more similar. There is evidence that \es\ becomes bluer at $t > +35$~d, although the data are noisy.

\begin{figure}[!t]
\begin{center}
\includegraphics[scale=.45]{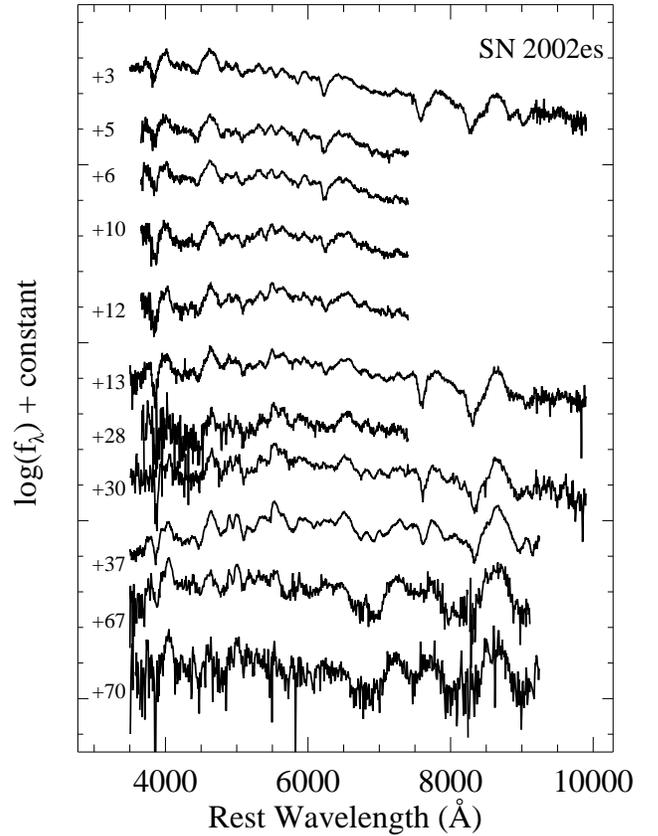}
\end{center}
{\caption{Montage of spectra of \es. All spectra have been shifted to the rest frame by the recession velocity of the host galaxy and corrected for $E(B-V)_{\rm MW} = 0.183$ mag. The phase relative to maximum light in the $B$ band is indicated to the left of each spectrum.} \label{f:spec_all}}
\end{figure}

\begin{figure}[!t]
\begin{center}
\includegraphics[scale=.43]{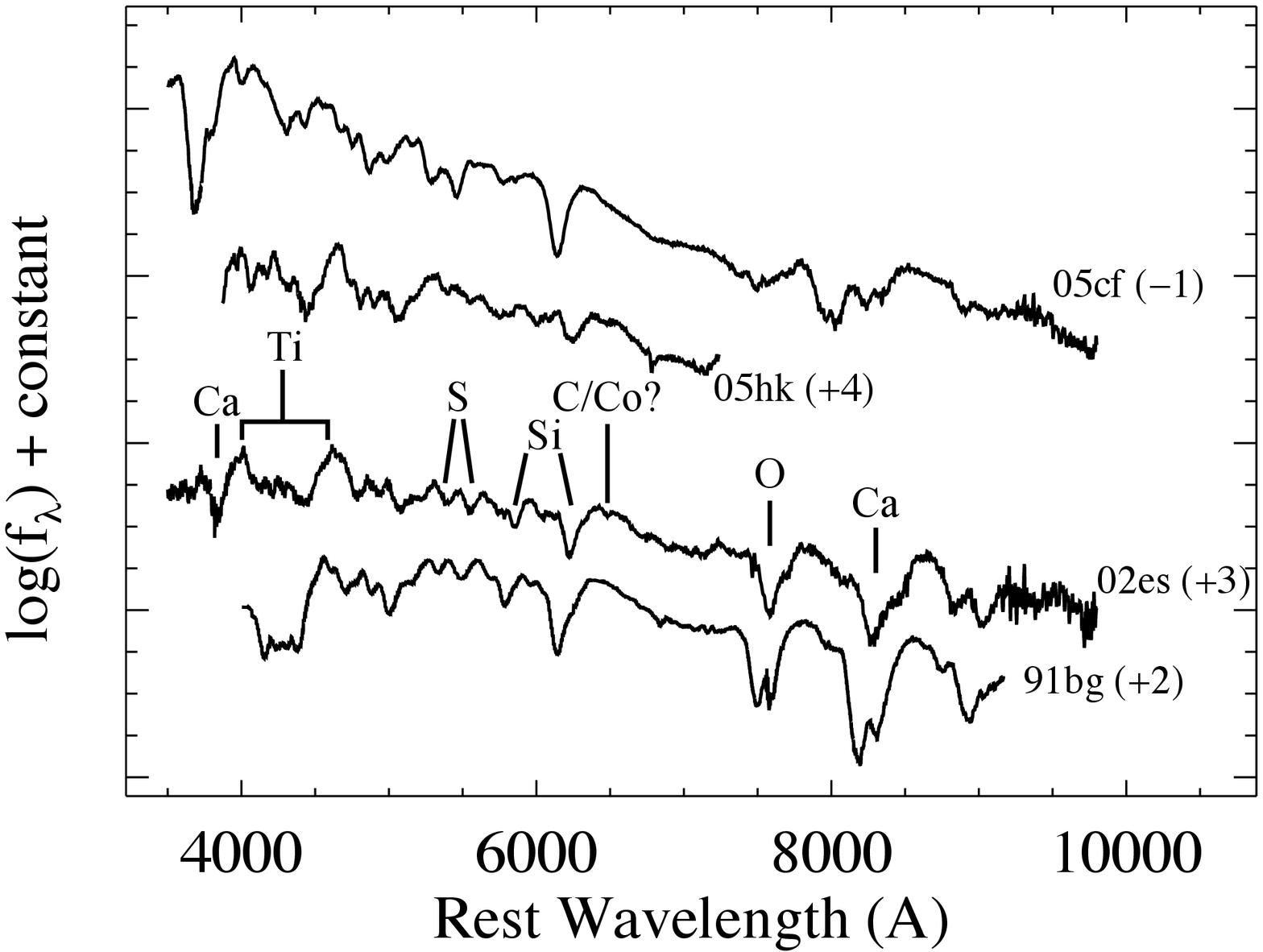}
\end{center}
{\caption{Spectrum of \es\ taken $+3$~d after maximum light in $B$, compared to the spectroscopically normal SN~Ia 2005cf \citep{wang09a}, the \cx-like \hk\ \citep{phillips07a}, and the subluminous \bg\ \citep{turatto96a} at similar phases. All objects have been shifted to the rest frame by their host-galaxy recession velocity and corrected for both Milky Way and host-galaxy extinction using the values in their respective references. Major spectroscopic features are identified. The spectrum of \es\ shares many similarities with \bg, except the former has significantly slower ejecta velocities. In particular, both objects show strong \ion{O}{1} and \ion{Ti}{2} features. The ejecta velocities of \es\ are more similar to those of \hk.} \label{f:early_comp}}
\end{figure}

\begin{figure}[!t]
\begin{center}
\includegraphics[scale=.43]{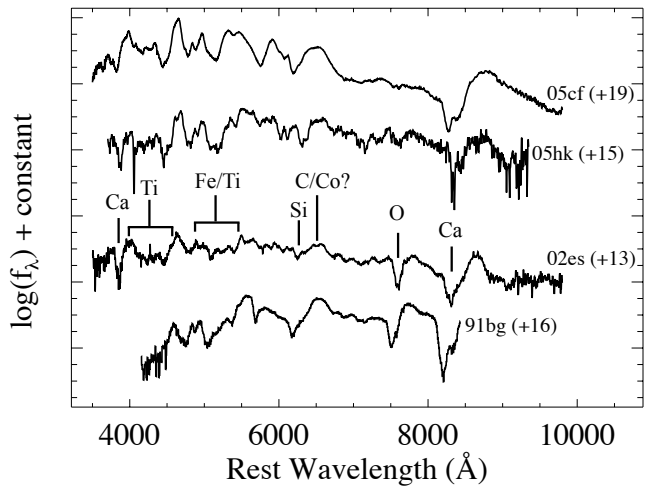}
\end{center}
{\caption{Spectrum of \es\ taken +13 d after maximum light. For comparison we show \cf\ \citep{wang09a}, \hk\ \citep{phillips07a}, and \bg\ \citep{turatto96a} at a similar phase indicated in parentheses to the right of each spectrum. All spectra have been corrected for both Milky Way and host-galaxy extinction using the values in their respective references. Major spectroscopic features are identified.} \label{f:spec_2wk_comp}}
\end{figure}

\begin{figure}[!t]
\begin{center}
\includegraphics[scale=.43]{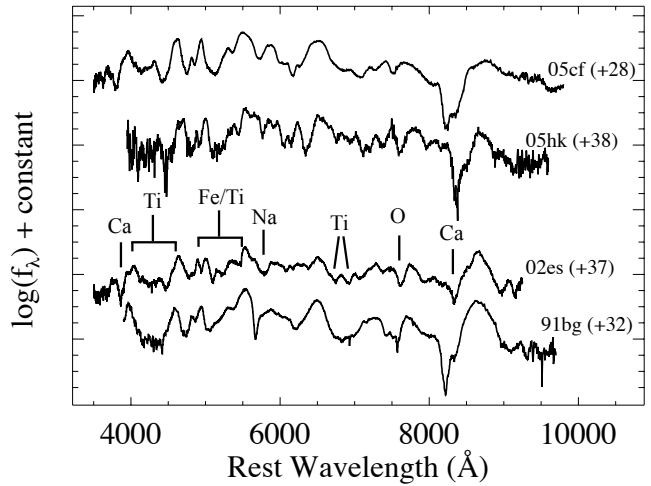}
\end{center}
{\caption{Spectrum of \es\ taken +37 d after maximum light using LRIS on Keck. For comparison we show \cf\ \citep{wang09a}, \hk\ \citep{phillips07a}, and \bg\ \citep{turatto96a} at a similar phase indicated in parentheses. All spectra have been corrected for both Milky Way and host-galaxy extinction using the values in their respective references.  Major spectroscopic features are identified.} \label{f:spec_1mon_comp}}
\end{figure}

\subsection{Spectral Properties}
In Figure \ref{f:spec_all}, we show our time series of spectra covering the evolution of \es\ starting 3 d after maximum light in the $B$ band and extending out to two months after maximum light. The absence of \hal\ and the presence of \ion{Si}{2} $\lambda$6355 in early-time spectra identify \es\ as a SN~Ia \citep[e.g.,][]{filippenko97a}. Strong \ion{Ti}{2} absorption around 4200~\AA\ is commonly associated with the subluminous \bg-like subclass of SNe~Ia \citep{filippenko92a}.

In Figure \ref{f:early_comp}, we show our earliest spectrum of \es\  compared to spectra of the normal \cf\ \citep{wang09a}, the subluminous \bg\ \citep[via the online SUSPECT database\footnote{http://suspect.nhn.ou.edu/$\sim$suspect .}]{turatto96a}, and the \cx-like \hk\ \citep{phillips07a}. At this phase, \es\ shares the most similarities with \bg. However, the expansion velocity of the photosphere as measured from the minimum in the blueshift of the P-Cyngni profile of \ion{Si}{2} $\lambda$6355 is 6000 \kms, significantly lower than typical values of \about 11,000 \kms\ for normal SNe~Ia \citep{wang09b,foley11b} and \bg-like SNe of \about 10,000 \kms\ \citep{taubenberger08a}. \cx-like objects have low ejecta velocities of \about 6000 \kms\   \citep{li03a,phillips07a}, comparable to what is found in \es. 

The spectrum of \hk\ only covers 3600--7400~\AA. Within this spectral range, it shows rather similar features to those of \es, but with notable differences as well. In particular, it has a weaker \ion{Si}{2} $\lambda$6355 absorption feature, and no obvious \ion{Ti}{2} absorption trough near 4200~\AA.

About two weeks after maximum light, \es\ begins to show more similarities with \cx-like objects. In Figure \ref{f:spec_2wk_comp}, we show our spectrum of \es\ from 13 d after maximum light compared to those of other objects at a similar phase. At this phase, \ion{Si}{2} $\lambda$6355 absorption is harder to discern. The \ion{Si}{2}/ \ion{Na}{1} complex around 5700~\AA\ is weaker than what is seen in \bg. Similar to \cx-like objects, \ion{Fe}{2} $\lambda\lambda$4555, 5129 absorption lines become more apparent and the \ion{Ca}{2} H\&K lines are similar in strength. However, \ion{O}{1} $\lambda$7774 and the \ion{Ca}{2} near-infrared (NIR) triplet are stronger in \es\ compared to \hk.

In Figure \ref{f:spec_1mon_comp}, we show our +37 d Keck spectrum along with comparison objects. We begin to see emission dominate in the case of the NIR \ion{Ca}{2} triplet. The low expansion velocities in \es\ and \hk\ make it possible to see narrow features which are usually hard to discern due to the smearing of broad high-velocity features in normal SNe~Ia. The spectrum is dominated by permitted iron-group elements, but line blending makes it difficult to uniquely identify features.

\begin{figure}[!t]
\begin{center}
\includegraphics[scale=.43]{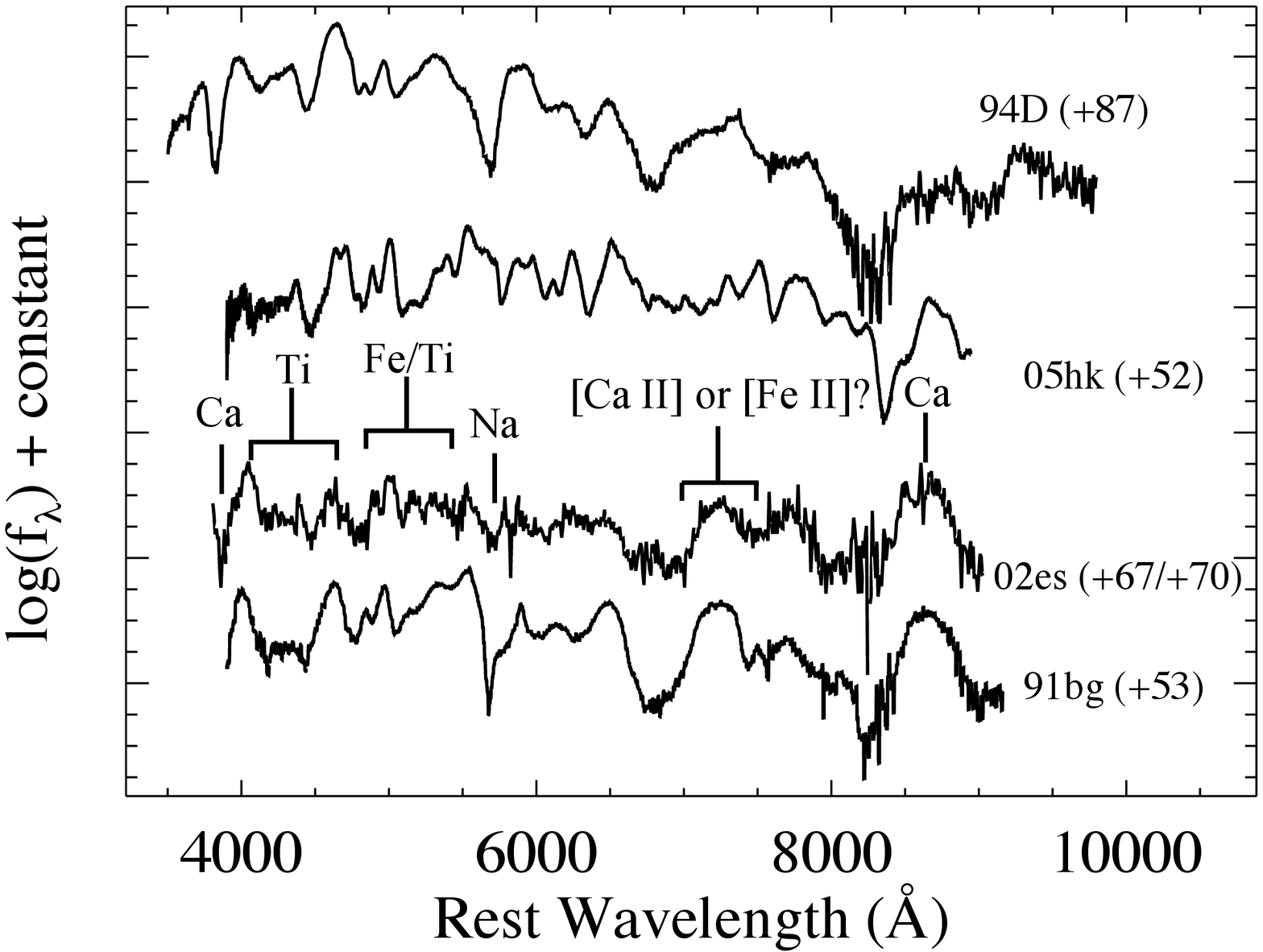}
\end{center}
{\caption{Combined spectrum of \es\ from +67 d and +70 d after maximum light. For comparison we show the normal SN 1994D (from our database of spectra),  \hk\ \citep{phillips07a}, and \bg\ \citep{turatto96a} at a similar phase indicated in parentheses to the right of each spectrum. All spectra have been corrected for both Milky Way and host-galaxy extinction using the values in their respective references. At this phase, \es\ looks similar to \bg, but with narrower features. Major spectroscopic features are identified.} \label{f:spec_2mon_comp}}
\end{figure}

We show our combined +67 d and +70 d spectrum  of \es\ in Figure \ref{f:spec_2mon_comp}, along with spectra of the normal SN~1994D, \hk, and \bg\ at comparable phases. \es\ continues to share the most similarities with \bg. Despite the rapid evolution of the light curves, the spectra do not show signs of being completely nebular. We continue to see continuum emission and absorption in our combined spectrum. There is a hint of forbidden emission lines, possibly [\ion{Ca}{2}] or [\ion{Fe}{2}] in the region around 7200 \AA, but we do not detect other prominent forbidden iron emission features commonly seen in nebular SN~Ia spectra \citep{stanishev07a,leloudas09a}. We note that the permitted \ion{Ca}{2} emission is not particularly strong, unlike PTF~09dav and the class of Ca-rich objects \citep{kasliwal11a}.

\begin{figure}[!t]
\begin{center}
\includegraphics[scale=.4]{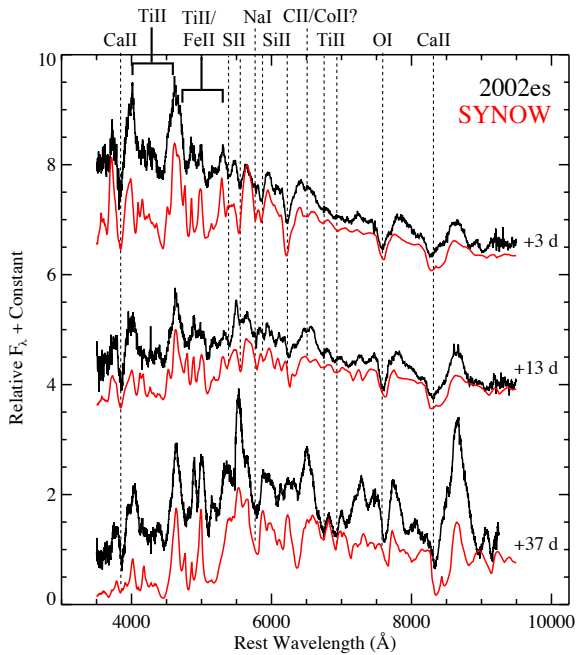}
\end{center}
{\caption{Spectra of \es\ at +3~d, + 13~d, and +37~d after maximum light compared to our best SYNOW fit for each epoch in red. Ions that contribute to major features are labeled according to the blueshifted absorption minimum of that feature. Our tentative identification of \ion{C}{2} is indicated by a question mark.} \label{f:sn2002es_synow}}
\end{figure}

\subsection{SYNOW Modeling \label{ss:synow}}
The supernova spectral synthesis code SYNOW \citep{fisher97a} is useful in identifying the different ion contributions in SN spectra. We use SYNOW as a tool to analyze our +3 d, +13 d, and +37 d spectra to determine the elemental composition of the photosphere. Our synthetic spectra compared to our observed spectra, along with identification of major spectral features. can be found in Figure \ref{f:sn2002es_synow}.

For our +3 d spectrum, we set the blackbody temperature to 13,000 K with a photospheric velocity of 6200 \kms. Many of the ions present in our SYNOW spectrum are commonly found in the spectra of \bg-like objects: \ion{Ca}{2}, \ion{Ti}{2}, \ion{O}{1}, \ion{Si}{2}, \ion{S}{2}, and \ion{Na}{1}.  \cite{sullivan11a} find evidence of \ion{Sc}{2} in spectra of PTF~09dav, a peculiar subluminous SN Ia similar in some characteristics to \es. We find that including \ion{Sc}{2} does not improve the fit to the spectrum of \es.

Just redward of \ion{Si}{2} $\lambda$6355 is a notch that looks suspiciously like \ion{C}{2} $\lambda$6580. The presence of carbon in pre-maximum spectra has recently been found in a substantial fraction (\about 20-30\%) of SNe~Ia \citep{parrent11a,thomas11a,folatelli12a} at velocities slightly higher than those of intermediate-mass elements such as Si and Ca. If the notch is due solely to \ion{C}{2}, it has a velocity of 4200 \kms, much lower than the velocities of other ions. With SYNOW, we can fit this feature reasonably well with a blend of \ion{C}{2} and \ion{Co}{2} with a velocity of 6200 \kms. We do not see evidence of \ion{C}{2} $\lambda$7234. The notch is also present in our spectrum at +13 d, but it is not generally seen in normal SNe~Ia that exhibit carbon at early times before maximum light. Persistent carbon features lasting two weeks were seen in the exceptionally luminous SN~2009dc \citep{taubenberger11a}. However, given the subluminous nature of \es, it shares few characteristics in common with SN~2009dc \citep{yamanaka09a, silverman11a,taubenberger11a} and similar overly luminous SNe~Ia \citep{howell01a,hicken07a,scalzo10a}. 

In an analysis of the pre-maximum spectrum of SN~2006bt, \cite{foley10b} found possible evidence for \ion{C}{2} at slower velocities (\about 5200 \kms ) than other ions (\about 12,500 \kms). However, unlike the case in \es, they found that the feature disappeared after maximum light. This feature, whether attributed to \ion{C}{2} or not, hints at a possible connection between the two transients.

In our +13 d spectrum, we set the blackbody temperature to 7500~K with a photospheric velocity of 5500 \kms. The low velocities of the ions allow for the detection of a forest of absorption lines normally smeared out by higher velocities. We see less evidence for the presence of \ion{Si}{2} and \ion{S}{2}. Absorption from iron-group elements such as \ion{Fe}{2} and \ion{Ti}{2} becomes more prominent. 

By +37 d, we use a blackbody temperature of 5900 K with a photospheric velocity of 3500 \kms. At this phase, we begin to see the SN become nebular and significant line blending at short wavelengths from iron-group elements. We still detect continuum emission, indicating that the SN is not completely nebular, but we begin to see emission from calcium.

\begin{figure}[!t]
\begin{center}
\includegraphics[scale=.55]{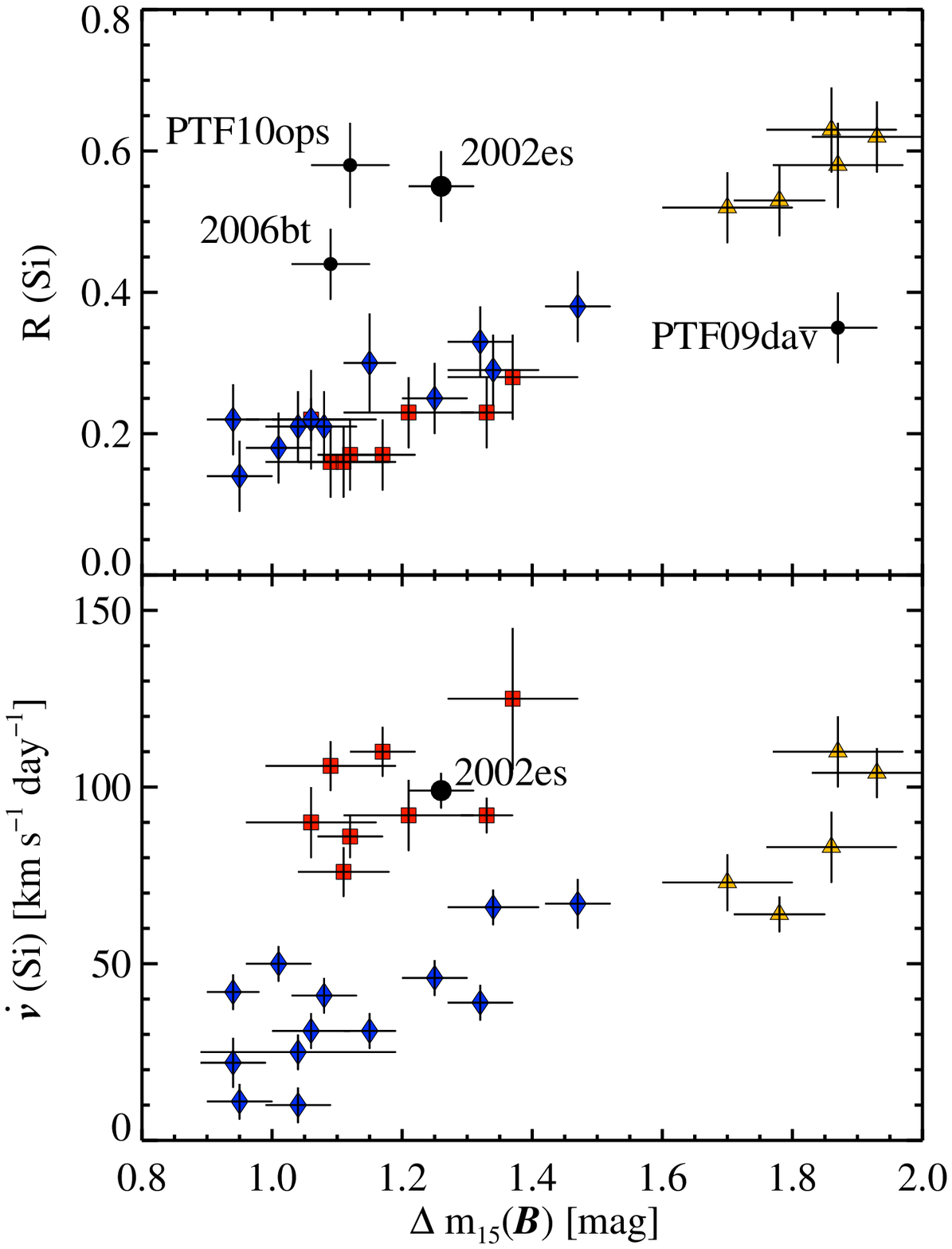}
\end{center}
{\caption{{\it Top panel:} The silicon ratio near maximum light, $\mathcal R$(\ion{Si}{2}), versus  \dmb. Blue diamonds are low-velocity objects , red squares are high-velocity objects, and yellow triangles are faint objects as defined by \cite{benetti05a}. \es\ joins PTF~10ops and SN~2006bt as outliers in the relationship between $\mathcal R$(\ion{Si}{2}) and \dmb. \es\ has a $\mathcal R$(\ion{Si}{2}) more typical of a faint object, despite having a broad light curve. {\it Bottom panel:} Gradient in the velocity of the \ion{Si}{2} feature versus \dmb.  Based on the clustering of this plot, \es\ would be classified as a HVG object despite being subluminous. } \label{f:sn2002es_benetti}}
\end{figure}

\subsection{Quantitative Measurements}
SNe~Ia can be broadly classified by a clustering analysis of spectral and photometric features \citep{benetti05a,branch06a}. Most quantitative measurements make use of \ion{Si}{2} features at 5972~\AA\ and \ion{Si}{2} $\lambda$6355. \cite{nugent95b} introduced $\mathcal{R} ({\rm Si})$, the ratio of the depth of \ion{Si}{2} $\lambda$5972 to $\lambda$6355 near maximum light, which is found to correlate with light-curve width. Using  $\mathcal{R} ({\rm Si})$ and the velocity gradient of the  \ion{Si}{2} $\lambda$6355 feature, $\dot{v}$, \cite{benetti05a} found that SNe~Ia could be broken into three distinct subclasses: low velocity gradient (LVG), high velocity gradient (HVG), and FAINT.

We estimate  $\mathcal{R} ({\rm Si})=0.55 \pm 0.05$ from our earliest spectrum of \es\ taken +3 d after maximum light. In the top panel of Figure \ref{f:sn2002es_benetti}, we plot \dmb\ against $\mathcal{R} ({\rm Si})$ for the objects in \cite{benetti05a} along with \es\ and other peculiar subluminous objects. While \es\ has a $\mathcal{R} ({\rm Si})$ value similar to that of FAINT objects, its measured \dmb\ does not match those of FAINT objects. Compared to other peculiar subluminous SNe~Ia, \es\ is most similar to PTF~10ops \citep{maguire11a}, both of which are significant outliers in the relationship between \dmb\ and $\mathcal{R} ({\rm Si})$ found by \cite{benetti05a}.

We measure $\dot{v} = 98 \pm 5~{\rm km~s^{-1}~d^{-1}}$ (statistical error only) from 6 spectra taken in the range $+3 \leq t \leq +13$~d. In the bottom panel of Figure \ref{f:sn2002es_benetti}, we plot \dmb\ against $\dot{v}$ for \es\ and the objects in \cite{benetti05a}, coded by subclass. Based on our measured $\dot{v}$ and \dmb, \es\ falls into the HVG group. FAINT objects also have high velocity gradients, but they also have a larger \dmb\ (i.e., a narrow light curve). \es\ would fall into the FAINT classification if it had a narrower light curve. Both panels indicate that \es\ has a unique combination of photometric and spectroscopic properties that lie outside the classification scheme of \cite{benetti05a}.

\begin{figure}[!t]
\begin{center}
\includegraphics[scale=.55]{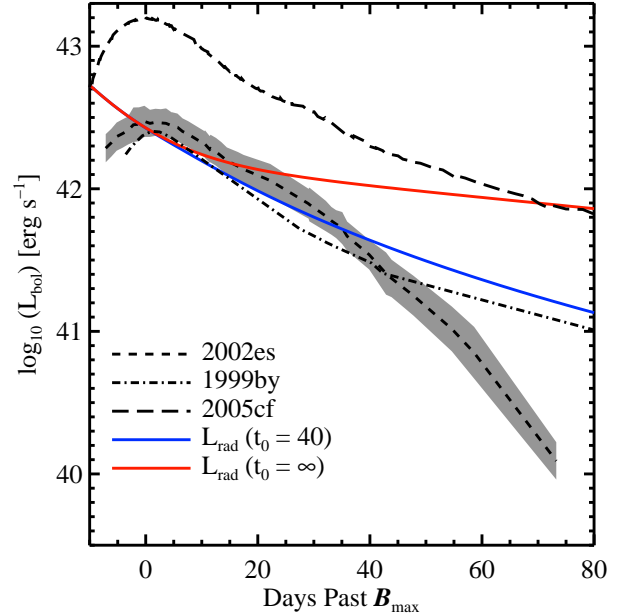}
\end{center}
{\caption{Bolometric luminosity of SN~2002es (dashed line) constructed using \bvri\ data and bolometric corrections from \bg. The shaded region indicates the $1\sigma$ errors. Plotted for comparison is the bolometric luminosity of SN~2005cf \citep[longer dashed line; ][]{wang09a} and SN~1999by \citep[dot-dash line; ][]{garnavich04a}. The solid blue and dotted red curves are models for the radioactive luminosity using Equation \ref{e:l_rad} assuming different $\gamma$-ray trapping efficiencies with an initial $M_{\rm Ni} = 0.17~{\rm M}_{\odot}$. The red dotted curve assumes full trapping of $\gamma$-rays producing a late-time slope that follows the cobalt decay rate. The blue curve is our best fit. Neither curve can reproduce the steep late-time decay in the bolometric light curve of \es.} \label{f:sn2002es_bol_lc}}
\end{figure}


\section{Discussion}
 \label{s:discussion}
\subsection{Bolometric Luminosity \label{ss:bol}}
Estimating the bolometric luminosity is particularly difficult without observations in the ultraviolet (UV) and IR, which contribute a sizable fraction of energy for normal SNe~Ia. Unfortunately, $U$-band photometry from \cite{hicken09a} starts \about 5 days after $B$-band maximum light. We only make use of \bvri\ data to construct a ``quasi-bolometric'' \citep{nomoto90a} luminosity light curve for \es. 

We estimate the quasi-bolometric luminosity  by warping spectra representing the SED of \es\  corrected for Milky Way extinction to match our multi-color photometry using a method similar to that of \cite{howell09a}. We assume there is no host-galaxy extinction (see \S \ref{ss:host}). For each epoch of \bvri\ photometry, we warp the spectrum using a third-order spline with knots at the effective wavelength of each filter  to match the photometric colors. In instances where we are missing photometry from one band, we do a linear interpolation between the nearest photometric epochs. We note that there is a fairly large gap in our $B$-band photometry at $+40 < t < +70$ d, making our interpolation uncertain. The resulting warped spectrum is integrated over the range 4000--8800~\AA\ (i.e., from the blue limit of the $B$ band to the red limit of the $I$ band) to obtain the optical flux for the photometric epoch. The flux is converted to a luminosity using the distance reported in \S \ref{ss:host}.

Ideally, a spectral series of \es\ matched with each photometric epoch (as opposed to a single spectrum) should be used for the most accurate results. The spectra from FAST presented in this paper do not extend to the $I$ band, and thus are not useful for estimating the bolometric luminosity. Given the spectroscopic similarities between \bg\ and \es, we used the above procedure with the \bg\  spectral series of \cite{nugent02a} artificially redshifted by 0.01 to match the slower expansion velocities of \es. 

A bolometric correction is required to turn our quasi-bolometric luminosity based on optical data into a bolometric luminosity that accounts for energy emitted in the UV and IR.  Our \bg\ spectral series covers 1000--25,000~\AA\ and provides a reasonable first-order approximation of the SED of \es. We calculate the fraction of flux emitted in the range 4000--8800~\AA\ compared to the total integrated flux for each spectrum in the series. This provides us with an estimate of the flux missed by only using optical data to construct our quasi-bolometric light curve as a function of phase. We apply this correction to arrive at our final bolometric light curve. At maximum light, the \bvri\ data account for \about70\% of the total flux. Given the uncertainties in our estimates, we include a 20\% systematic error in our error budget for the bolometric luminosity.

We estimate $L_{\rm bol} =  (4.0 \pm 0.9) \times 10^{42}~\rm{erg~s^{-1}}$ at \about 1~d after maximum light in the $B$ band. This is slightly larger than the bolometric luminosity of other subluminous SNe~Ia \citep{taubenberger08a} and \about 25\% of the luminosity of  \cf\ \citep{wang09a}.

The luminosity of thermonuclear SNe is powered by the energy deposition of $\gamma$-rays and positions produced by the radioactive decay chain $^{56}$Ni $\to$ $^{56}$Co $\to$ $^{56}$Fe. At maximum light, the rate of energy deposition into the expanding ejecta is roughly equivalent to the luminosity of the SN \citep[i.e., Arnett's law; ][]{arnett82a}. Following \cite{stritzinger05b}, we can write Arnett's law as

\begin{eqnarray}\label{e:arnett}
L_{\rm bol} = \alpha \times (6.45e^{{-t_r}/(8.8{\rm d})} + 1.45e^{{-t_r}/(111.3{\rm d})}) \times  \\
 \left(\frac{M_{\rm Ni}}{{\rm M}_{\odot}} \right) \times 10^{43}~ {\rm erg~s^{-1}}, \nonumber
\end{eqnarray}
where $\alpha$ is a correction factor of order unity to Arnett's law and $t_{r}$ is the time between explosion and maximum light (i.e., the bolometric rise time).

Assuming the luminosity of \es\ is powered by the decay of \nic, we can estimate the amount of \nic\ synthesized in the explosion.  We set $\alpha = 1$. Unfortunately, we do not have tight constraints on the date of explosion. Prior to the first detection of \es\ on 2002 Aug. 23.2 (\about 9~d before the time of maximum $L_{\rm bol}$), KAIT obtained an unfiltered image of the field on 2002 Aug. 12.5 with a limiting magnitude of 18.5 \citep{li02a}. \cite{ganeshalingam11a} found that the subluminous \by\ had a rise time to maximum light in $B$ of $13.33 \pm 0.40$~d, consistent with other estimates for the rise time of \bg-like SNe from \citet{taubenberger08a}, while a typical, normal SN~Ia has a rise time of \about 18~d. The $B$-band light curve of \es\ has a slower rise to maximum light than \by\ and matches the rise of \hk. \cite{phillips07a} measure a rise to $B$-band maximum of $15 \pm 1$~d for the \cx-like \hk. \cite{maguire11a}  constrain the rise time of PTF~10ops, an object that shares similarities to \es, to \about19~d. Based on the unique spectroscopic and photometric peculiarities of \es, it is unclear which object serves as the best guide to determining the rise time. As a compromise between the different possible rise times, we adopt a rise time to bolometric maximum of $t_{r} = 16 \pm 3$~d.

From Equation \ref{e:arnett}, we estimate $M_{\rm Ni} = 0.17 \pm 0.05~{\rm M}_{\odot}$ synthesized in the explosion which falls at the low end of the range  $0.05<M_{\rm Ni}< 0.87~{\rm M}_{\odot}$ found by \cite{stritzinger06a} for a sample of SNe~Ia. \bg\ and \by\ synthesized \about 0.1 \Msun\ of \nic.

\subsection{Energetics \label{ss:energetics}}
We can estimate the ejected mass, $M_{\rm ej}$, and the kinetic energy, $E_{0}$, of the explosion using the rise time. Following the treatments of \cite{arnett82a} and \cite{pinto00b,pinto00a}, we have $t_{r}^{2} \propto \kappa M_{\rm ej} / v_{s}$, where $\kappa$ is the mean opacity and $v_{s}$ is the ejecta velocity. If we compare \es\ to a normal SN~Ia with $M_{\rm ej} \approx 1.4~{\rm M}_{\odot}$, $v_{s} = 10^{4}~{\rm km~s^{-1}}$, and $t_{d} = 18~{\rm d}$, and assume they have similar opacities, we have
\begin{equation}
M_{\rm ej} = 0.66 {\rm M}_{\odot} \left(\frac{t_{r}}{16~{\rm d}}\right )^{2} \left( \frac{v_{s}}{6000~{\rm km~s^{-1}}}\right).
\end{equation}
The largest uncertainty is our calculation is from our estimate of the rise time of $t_{r} = 16 \pm 3$~d. Including this uncertainty, we find $M_{\rm ej} = 0.66 \pm 0.25$ for our nominal values, significantly lower than the canonical values for a SN Ia.

We can then calculate the kinetic energy
\begin{equation}
E_{\rm k} = 2.4 \times 10^{50}~{\rm erg}  \left (\frac{t_{r}}{16~{\rm d}}\right )^{2} \left ( \frac{v_{s}}{6000~{\rm km~s^{-1}}}\right)^{3}.
\end{equation}
Including the uncertainty in our rise time gives $E_{\rm k} = (2.4 \pm 0.9) \times 10^{50}~{\rm erg}$ for our nominal values.

The estimated ejected mass and kinetic energy are significantly lower than the canonical values for a SN~Ia in part due to the slow expansion velocities measured from the Si feature. We again caution that our estimate for the rise time is not well constrained and can range from 13~d to 19~d. 

Following similar arguments made by \cite{howell06a} and \cite{silverman11a}, we can place constraints on the white dwarf (WD) progenitor mass, $M_{\rm WD}$, based on the energetics of the explosion. The amount of energy produced by burning carbon and oxygen up to intermediate-mass elements (IMEs) and iron-group elements (IGEs) must equal the energy required to unbind the WD and the kinetic energy of the ejecta: $E_{\rm nuc} = E_{\rm k} + E_{\rm binding}$.  The energy released from fusing a WD consisting of equal parts carbon and oxygen to iron-group elements is $1.55 \times 10^{51}~{\rm erg}~{\rm M}_{\odot}^{-1}$, and fusion up to $^{28}{\rm Si}$ releases 76\% as much energy \citep{branch92a}. Assuming the entire WD is burned in the thermonuclear explosion, the fraction of iron-group elements, $f_{\rm IGE}$, and the fraction of intermediate-mass elements, $f_{\rm IME}$, will add to 1. The energy released from nuclear fusion can be written as 
\begin{equation}
E_{\rm nuc} = 1.55 ~\times 10^{51}~{\rm erg} \left ( \frac{M_{\rm WD}}{{\rm M}_{\odot}} \right ) (f_{\rm IGE} + 0.76 f_{\rm IME}).
\end{equation}
The kinetic energy is given by 
\begin{equation}
E_{\rm k} = \frac{1}{2} M_{\rm WD}v^{2} = 3.6 \times 10^{50}~{\rm erg~s^{-1}} \left ( \frac{M_{\rm WD}}{{\rm M}_{\odot}} \right ),
\end{equation}
where we have set the $v = 6000$~\kms. For the binding energy, we construct a series of WD models with central temperature $T = 10^{7}$~K using the Models for Experiments in Stellar Astrophysics (MESA) code \citep{paxton11a}. We then calculate the binding energy as a function of $M_{\rm WD}$ within the range $0.7 < M_{\rm WD} < 1.4~{\rm M}_{\odot}$ (i.e., the expected range of masses for a sub-Chandrasekhar double-detonation model).  Models of sub-Chandrasekhar-mass WD explosions generally find that the bulk of IGEs synthesized is in the form of \nic\ \citep{sim10a, woosley11a}. Based on our calculated \nic\ mass in \S \ref{ss:bol}, we then have 
\begin{eqnarray}
f_{\rm IGE} &=& 0.17 \left (\frac{M_{\rm WD}}{{\rm M}_{\odot}} \right )^{-1}~{\rm and} \\
f_{\rm IME} &=& 1 - f_{\rm IGE}.
\end{eqnarray}

Over the range of WD masses, we find that the nuclear energy released is larger than the combined binding energy and kinetic energy. This ensures that there is enough energy to unbind the WD and power the explosion based on the \nic\ mass, but it is unclear what happens to the excess nuclear energy. Realistically, some of the nuclear energy should go into heating the ejecta. If some fraction of the WD was left unburned, then $E_{\rm nuc}$ would decrease. However, our evidence for unburned carbon (see \S \ref{ss:synow}) is questionable. Given the uncertainty in our assumptions, we cannot place a strong constraint on the mass of the progenitor.

\subsection{Late-Time Decay \label{ss:late}}
We build a simple toy model to describe the luminosity due to the deposition of energy from the radioactive decay of \nic\ and \cob\ to study the behavior of the bolometric light curve. The energy initially deposited into the ejecta is from the thermalization of $\gamma$-rays emitted by radioactive decay of $^{56}{\rm Ni}\! \to ^{56}{\rm Co}$ with an $e$-folding time of 8.8~d. By \about\ 20~d, the energy emitted by the decay of $^{56}{\rm Co} \! \to ^{56}{\rm Fe}$ exceeds that of \nic\ ($e$-folding time of 111.3 d). The decay of \cob\ can proceed via either electron capture emitting a spectrum of $\gamma$-rays (96.5\% of emitted energy) or beta decay releasing a positron ($3.5\%$ of emitted energy) \citep{nadyozhin94a}. The luminosity due to the radioactive decay can be written as
\begin{eqnarray}
L_{\rm rad} &  = &  L_{\rm Ni,\gamma} [1 - e^{-\tau}]  +     \\
  &   &  L_{\rm Co, e^{+}}  + L_{\rm Co, \gamma} [1 - e^{-\tau}], \nonumber
\end{eqnarray}
where the factor $(1 - e^{-\tau})$ is the fraction of $\gamma$-rays trapped in the expanding ejecta with an optical depth $\tau$. We assume that the positrons are fully trapped within the ejecta and deposit their kinetic energy instantaneously. Note that $\tau$ is a function of time, decreasing as the ejecta expand homologously. Following \cite {sollerman02a,sollerman04a} and \cite {stritzinger06b}, we write the optical depth to $\gamma$-rays in expanding ejecta as
\begin{equation}
\tau = \left(\frac{t_{0}}{t} \right)^{2},
\end{equation}
where $t_{0}$ is the fiducial time relative to explosion when the ejecta become optically thin to $\gamma$-rays.

Solving the first-order differential equations that describe the parent-daughter relationship between \nic\ and \cob, we write the radioactive luminosity as
\begin{eqnarray} \label{e:l_rad}
L_{\rm rad}  &=& \lambda_{\rm Ni} Q_{\rm Ni, \gamma} N_{\rm Ni, 0} {e ^{- \lambda_{\rm Ni} t}} [1 - e^{-\tau}]  \\
  &  &   + \frac{\lambda_{\rm Ni} \lambda_{Co}}{\lambda_{\rm Ni} - \lambda_{\rm Co}} N_{\rm Ni,0}(e^{-\lambda_{\rm Co}t} - e^{-\lambda_{\rm Ni}t} ) \nonumber \\
 & & \times \left [ Q_{\rm Co, e^{+}} + Q_{\rm Co, \gamma}(1 - e^{-\tau}) \right ], \nonumber
\end{eqnarray}
where $\lambda_{\rm Ni}$ and $\lambda_{\rm Co}$ are the inverse $e$-folding times for \nic\ and \cob, respectively; $N_{\rm Ni,0}$ is the initial amount of \nic\ synthesized in the explosion; $Q_{\rm Ni, \gamma}$(1.75 MeV) is the energy yielded by each $^{56}{\rm Ni} \to ^{56}{\rm \!\!\!Co}$ decay; and $Q_{\rm Co, \gamma}$(3.61 MeV) and $Q_{\rm Co,e^{+}}$(0.12 MeV) are the energy yielded per $^{56}{\rm Co} \to ^{56}{\rm \!\!Fe}$ decay via electron capture and beta decay, respectively.

In Figure \ref{f:sn2002es_bol_lc}, we plot the bolometric light curve of \es\ (see \S \ref{ss:bol} for details), as well as the UVOIR bolometric light curves of \cf\ from \cite{wang09a} and SN~1999by using data from \cite{garnavich04a}. We also plot models $L_{\rm rad}(t_{0} = 40~{\rm d})$ and $L_{\rm rad} (t_{0} = \infty~{\rm d})$ assuming $M_{\rm Ni} = 0.17~{\rm M}_{\odot}$ (found using Arnett's law in \S \ref{ss:bol}). The model with $t_{0} = \infty$ represents the case in which the $\gamma$-rays are completely trapped and the luminosity decays according to the \cob\ decay rate (0.01 \md).

The bolometric light curve of \es\ shows a surprising drop in luminosity at $t > +30$~d. The decline is significantly faster than the cobalt decay rate and the decline rate for \cf.  Similar results were found for the late-time decline in individual optical bands (see \S \ref{s:phot}). Even when accounting for a $\gamma$-ray optical depth that decreases with time, we are unable to find a model that reasonably matches the bolometric light-curve decline rate of \es. Our best match is $t_{0} = 40$~d, which adequately describes \es\ within the region $0 < t < +30$~d but fades too slowly at $t > +30$~d.

The decay of the bolometric light curve in the range $+30 < t < +70$~d appears incompatible with the gradual escape of $\gamma$-rays through homologously expanding ejecta and implies that the ejecta become optically thin to $\gamma$-rays very rapidly. If the ejecta are optically thin to $\gamma$-rays, then the luminosity should be powered by the thermalized kinetic energy of positrons, assuming some fraction of the positrons are trapped in the ejecta. We estimate the nickel mass required to power the light curve by positrons at our last photometry epoch ($t = 89.3 \pm 3$~d after explosion) using Equation \ref{e:l_rad} and setting $ \tau = 0$ (i.e., the ejecta are optically thin to $\gamma$-rays).  For complete positron trapping, we estimate  $M_{\rm Ni} = 0.05 \pm 0.02~{\rm M}_{\odot}$ is required to power the luminosity at this phase; this is a factor of 3 less than our previous estimate using Arnett's law at maximum light. A corollary to the assumption of complete positron trapping is that the light-curve decay should follow the cobalt decay rate, which is clearly not seen. If we allow for only partial trapping, we need a positron trapping fraction of \about 0.3 to match our $M_{\rm Ni}$ from Arnett's law.

The unexpectedly fast decline calls into question whether \es\ was powered by \nic\ decay. Here we discuss possible ways to reconcile a thermonuclear SN with the measured decline rate.

Dust formation around the SN would lead to a drop in optical flux as high-energy photons are reprocessed to longer IR wavelengths. In models of dust formation in the ejecta of SNe~Ia, \citet{nozawa11a} find that the conditions required to produce dust occur 100--300 d after maximum light. However, due to the low densities in SN~Ia ejecta, the dust grains are small ($ < 0.01~\mu{\rm m}$), and the expected IR emission associated with dust has not yet been detected in late-time observations of SNe~Ia \citep{gerardy07a}. We  also do not find evidence of an increase in the red continuum in our spectra or asymmetries in line features in our late-time spectra as have been seen in other SNe that formed dust \citep{smith08a}. Our last epoch of photometry taken +73 d after maximum light gives $B - R \approx 0.5$~mag, while photometry of \cf\ taken from \cite{wang09a} gives  $B-R \approx 0.7$ mag. \es\ is bluer than \cf, which is opposite the effect we would expect if dust were facilitating the rapid fading.

\cite{axelrod80a} predicts that SNe~Ia should undergo an ``infrared catastrophe" (IRC) at late times, once the temperature drops below a critical threshold (\about 1000 K). Models of the IRC predict that a thermal instability shifts the bulk of emission from the optical to fine-structure transitions of iron in the IR about 500--700 d after maximum light. Without IR data, we are unable to determine whether the IR flux increases as the optical flux decreases, although the onset of the IRC at such an early phase is certainly unexpected. Our last spectrum of \es\ taken 70 d after maximum light shows permitted lines in absorption, indicating that the ejecta have not yet become nebular and the temperature has probably not dropped sufficiently low to facilitate the onset  of the IRC. The IRC has never been detected in late-time observations of other SNe~Ia, although \cite{leloudas09a} evoke the possibility of the IRC occurring locally in clumpy ejecta to explain the missing flux in the late-time light curve of SN 2003hv.

\subsection{A Pure Explosion Model}
Another possible model for the evolution of the light curve of \es\ is that the energy deposited into the ejecta is derived from the explosion of the progenitor, and there is no subsequent heating. Such a model was explored by \cite{kasliwal10a} to explain the rapidly evolving SN~2010X. The explosion energy  ($E_{0}$) is deposited instantaneously into the ejecta. At maximum light,  $L_{\rm max} \approx E_{0}/t_{d}(0)$, where $t_{d}(0)$ is the initial diffusion time for photons through the ejecta. The diffusion time is $ t_{d}(0)  \propto M_{\rm ej} \kappa / R_{0}$, where $R_{0}$ is the initial radius of the SN. Assuming $\kappa = 0.1~{\rm cm^{2}~g^{-1}}$ for Fe-rich ejecta \citep{pinto00a},   $L_{\rm max} = 4.0 \times 10^{42}~{\rm erg~s^{-1}}$, and $M_{\rm ej} = 0.66~{\rm M}_{\odot}$,  we estimate an initial progenitor radius of $R_{0} \approx 10^{12}~\rm{cm}$. 

This value is similar to the estimate presented by \cite{kasliwal10a}, in which they argue that such a radius would require a progenitor with an extended hydrogen envelope. Based on the absence of hydrogen in spectra of SN~2010X they reject this hypothesis. Similarly, given the lack of hydrogen in our spectra of \es, we also find a pure explosion an unlikely mechanism to power the luminosity of \es.

\subsection{A Core-Collapse SN? \label{ss:cc}}
If \es\ was not powered by \nic\ decay, it is reasonable to investigate core-collapse mechanisms that could explain the properties of \es. However, the lack of hydrogen emission in optical spectra of \es\ rules out a massive star with a large hydrogen envelope as a progenitor, leaving the possibility that \es\ could be a SN~Ic. The star-formation history of the host galaxy strongly favors an older stellar population, making the progenitors of SNe~Ibc unlikely as well \citep{leaman11a}. If \es\ is a core-collapse event, it would have to be the result of a low-mass star following an atypical evolutionary path to a SN~Ic.

\cite{perets10a} discussed data on SN~2005E, a low-luminosity SN~Ib (based on the absence of hydrogen and the presence of helium) with calcium-rich ejecta that exploded in the outskirts of its early-type host galaxy. Although the observed properties of SN~2005E and \es\ are different (e.g., \es\ is much brighter, has slower expansion velocities, and no detectable helium), SN 2005E and other objects like it may be core-collapse SNe connected to old stellar populations. \cite{kasliwal11a} studied the observed properties of the emerging class of Ca-rich objects and found that they could not be explained by conventional core-collapse or thermonuclear explosions.

Another possible atypical core-collapse object was SN~2008ha, a low-luminosity Type I SN (based on the absence of hydrogen) that peaked at $-14.2$ mag with extremely low ejecta velocities (\about 4000--5000 \kms; \citealt{valenti09a,foley09a,foley10c}). \cite{valenti09a} argued that SN~2008ha and the family of \cx-like objects may be the result of the core collapse of hydrogen-poor, low-mass stars. Their analysis of SN~2008ha spectra showed little evidence for the IMEs commonly associated with the byproducts of a thermonuclear explosion and an absence of forbidden iron lines at late times. Coupled with the low luminosity and low ejecta velocities, the authors interpret SN~2008ha as a core-collapse event that produced little \nic. However, \cite{foley10c} presented an early-time spectrum of SN~2008ha that clearly exhibits IMEs such as silicon, sulfur, and carbon, arguing in favor of the thermonuclear explosion of a C/O WD.

\begin{deluxetable*}{ccccc}
 \tablewidth{0pc}
 \tablecaption{Log of Optical Spectral Observations of \bh\label{t:sn1999bh_spec}}
 \tablehead{\colhead{UT Date} & \colhead{Phase\tablenotemark{a} (d)} & \colhead{Telescope/Instrument}& \colhead{Exp. Time (s)}& \colhead{Observer\tablenotemark{b}}}
 \startdata
1999 Apr. 09.3   & +5  &  FLWO/FAST& 1200 & MC \\
1999 Apr. 24.4   & +20 &  Lick/Kast& 1800 & AF, WL
\enddata
\tablenotetext{a} {Rest-frame days relative to the date of $B_{\rm max}$, 1999 Apr. 03.3 (JD 2,451,271.8), rounded to the nearest day.}
\tablenotetext{b} {AF = A. Filippenko, MC = M. Calkins, WL = W. Li.}
\end{deluxetable*}

\begin{figure}[!t]
\begin{center}
\includegraphics[scale=.55]{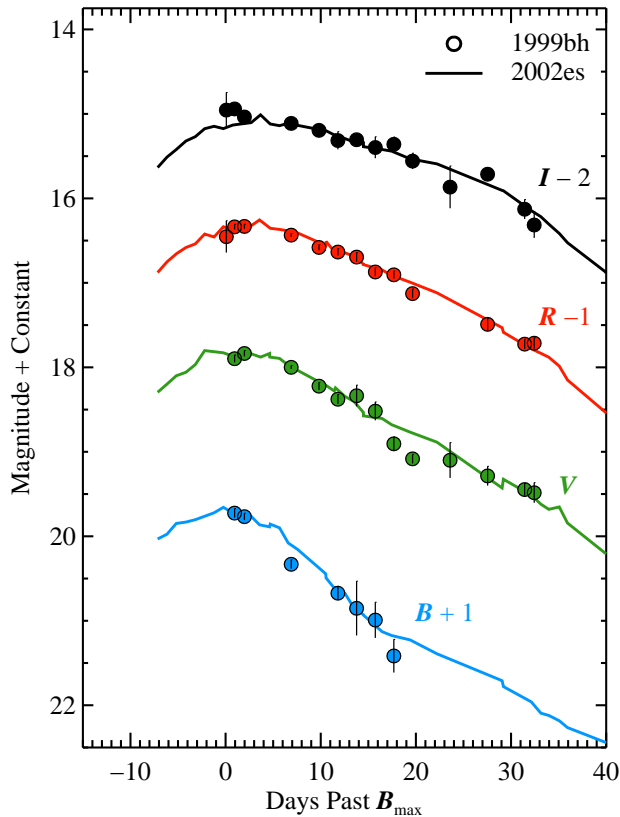}
\end{center}
{\caption{\bvri\ light curves of \bh\ (filled circles) in comparison to \es\ (solid lines). The light curves of both objects have been shifted relative to the time of $B_{\rm max}$ and peak magnitude.  The light curves of \bh\ display a striking similarity to those of  \es. Both sets of light curves are particularly broad, despite being subluminous and spectroscopically similar to those of \bg.} \label{f:sn1999bh_lc}}
\end{figure}

\begin{figure}[!t]
\begin{center}
\includegraphics[scale=.55]{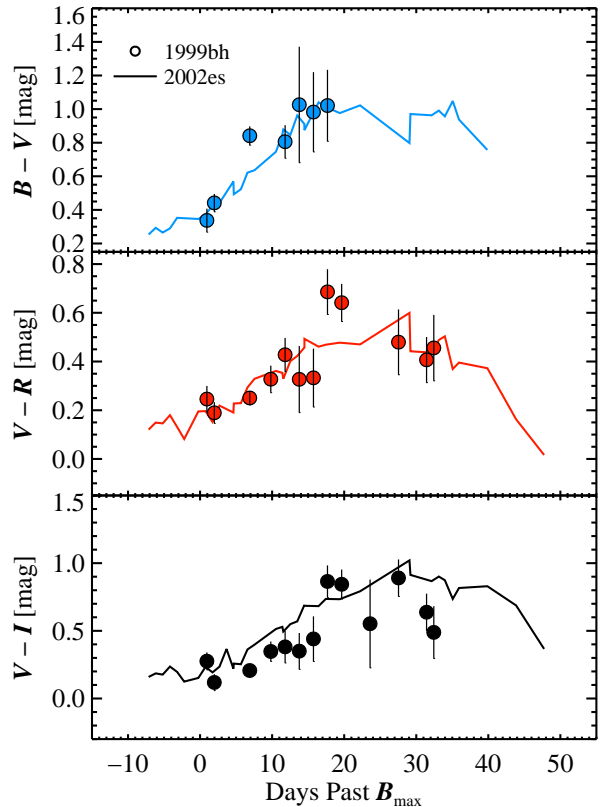}
\end{center}
{\caption{Color curves of \bh\ (solid circles) compared to \es\ (solid lines). \bh\ has been corrected for $E(B-V)_{\rm MW} = 0.015$ mag and $E(B-V)_{\rm host} = 0.48$~mag. The color curves of \es, corrected for $E(B-V)_{\rm MW} = 0.183$~mag, provide an excellent match for \bh.} \label{f:sn1999bh_col}}
\end{figure}

\begin{figure}[!t]
\begin{center}
\includegraphics[scale=.45]{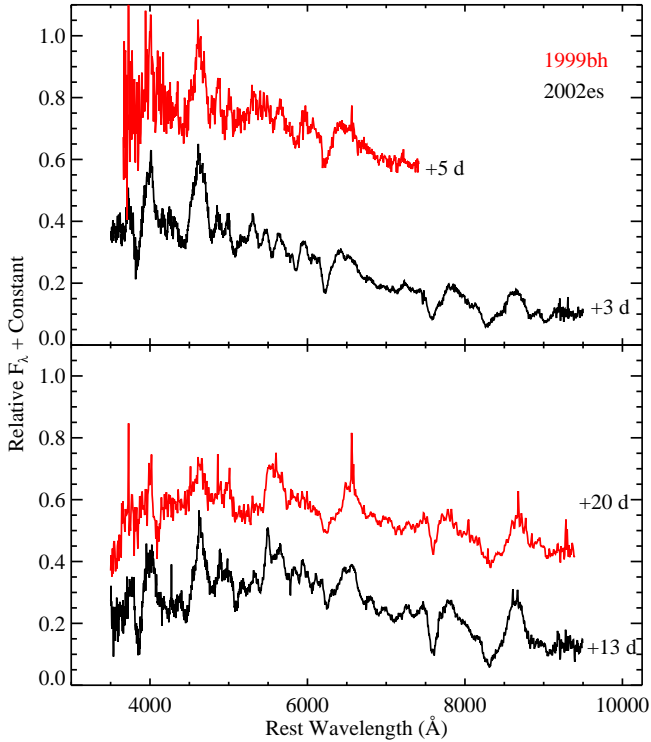}
\end{center}
{\caption{Spectra of \bh\  (red) in comparison to \es\ (black). The spectra have been corrected for the effects of host-galaxy recession. Spectra of  \bh\ have been corrected for $E(B - V) = 0.01$~mag due to Milky Way extinction and $E(B-V) = 0.48$~mag from host-galaxy extinction using a CCM reddening law with $R_{V} = 3.1$. Spectra of \es\ have been corrected for $E(B-V) = 0.183$~mag due to Milky Way extinction.  } \label{f:sn1999bh_spec}}
\end{figure}

\subsection{SN 1999bh: A SN~2002es-Like Object}
Combing through our photometric \citep{ganeshalingam10a} and spectral \citep{silverman12a} databases, we recognized that \bh\ shares many of the same properties as SN~2002es. Although our photometric and spectroscopic coverage of \bh\ is not as extensive as our \es\ dataset, the available data provide a compelling case to link the two objects.

$BV\!RI\!$ photometry of \bh\ are taken from \cite{ganeshalingam10a}. Spectra of the object were obtained using the Kast dual spectrograph on the Shane 3-m telescope at Lick Observatory and the FAST spectrograph mounted on FLWO at Mount Hopkins. A journal of our observations is available in Table \ref{t:sn1999bh_spec}. The spectra were reduced using the techniques described in \S \ref{ss:spec}.

SN~1999bh was discovered in NGC 3435 by \cite{li99a} as part of LOSS on 1999 March 29.2. Subsequent spectroscopic follow-up observations by \cite{aldering99a} classified the object as a SN~Ia near maximum light on 1999 Apr. 02. The authors estimate a redshift of 0.028, likely based on SN features. The redshift of NGC 3435 listed in NED is $z_{\rm helio} = 0.0172$. We measure a redshift of $z_{\rm helio} = 0.0168$ from narrow H$\alpha$ + [\ion{N}{2}] lines from the host galaxy in a spectrum of SN~1999bh taken on 1999 Apr. 24 using the Kast dual spectrograph mounted on the Shane 3-m telescope. We adopt this redshift for the remainder of our analysis. The luminosity distance to NGC 3435 calculated using $z_{\rm CMB} = 0.0172$ is $d_{L} = 70.8 \pm 5.3$~Mpc. At this redshift, SN~1999bh exploded at a projected distance of 3.5 kpc from the nucleus of NGC 3435.

Figure \ref{f:sn1999bh_lc} shows the \bvri\ light curves of \bh; we see that those of \es\ provide an excellent match in all bands. Both objects exhibit broad light curves despite being subluminous, and lack a prominent shoulder in the $R$ or $I$ bands.

Using \es\ as a template, we measure the date of maximum light to be JD $2,451,271.8 \pm 1$~d (1999 Apr. 3.6). After correcting for $E(B-V) = 0.015$~mag from Milky Way extinction using the dust maps of \cite{schlegel98a}, we measure $B_{\rm max} = 18.63 \pm 0.06$ mag. Due to the faintness of the object, we were unable to follow \bh\ sufficiently long to measure the light-curve decay rates at $t > +30$~d.

Based on the presence of \ion{Na}{1}~D absorption at the redshift of NGC 3435 in spectra of \bh, we can be certain there is some amount of extinction due to the host galaxy. We estimate an equivalent width of $0.8 \pm 0.2$~\AA\ of \ion{Na}{1}~D absorption at the redshift of the host galaxy from our spectrum taken on 1999 Apr. 09. Translating \ion{Na}{1}~D EW measurements to an inferred $E(B-V)$ reddening gives  $0.1 < E(B-V) < 0.2~{\rm mag}$ depending on whether we use the relationship given by \cite{barbon90a} or \cite{turatto03a}. \cite{blondin09a} and \cite{poznanski11a}, however, have shown that while there is a positive correlation between the presence of \ion{Na}{1}~D absorption and host-galaxy extinction, \ion{Na}{1}~D absorption is not a strong predictor for the amount of extinction.

Instead, we estimate the amount of host-galaxy extinction by matching the $B - V$ color of \bh\ to that of \es. This assumes that any difference in colors is purely associated with host-galaxy extinction and not intrinsic differences between the colors of the two objects. We find that $E(B-V)_{\rm host} = 0.48 \pm 0.07$~mag.

We also estimate the host-galaxy reddening by matching our \bh\ spectra to corresponding \es\ spectra. We perform a fit to match the SED of the two objects applying a CCM reddening law \citep{cardelli89a} with $R_{V} = 3.1$ to deredden our \bh\ spectra. We obtain a best fit of $E(B-V)_{\rm host} = 0.59 \pm 0.01$~mag (statistical error only) matching our +5 d spectrum of \bh\ to our +3 d spectrum of \es, and $E(B-V)_{\rm host} = 0.49 \pm 0.01$~mag matching our +20 d spectrum of \bh\ to our +13 d spectrum of \es. These values are consistent with what we derived above using the $B-V$ color at maximum light. We adopt $E(B-V)_{\rm host} = 0.48 \pm 0.07$~mag as the host-galaxy extinction.

In Figure \ref{f:sn1999bh_col} we show the color curves of \bh\ corrected for Milky Way and host-galaxy reddening. The curves of \es\ provide an excellent match for the color evolution of \bh. The $B-V$ and $V-R$ colors agree almost perfectly. The $V-I$ color of \bh\ is bluer than expected, compared to that of \es.

After correcting for Milky Way and host-galaxy extinction, the absolute magnitude at peak of \bh\ was $M_{B} = -17.71 \pm 0.27$~mag, almost the same luminosity as \es. We estimate $\Delta m_{15}(B) = 1.24 \pm 0.10$~mag. Having a light-curve width that is comparable to that of \es, \bh\ is another example of a $5\sigma$ outlier in the Phillips relation as shown in Figure \ref{f:phillips}.

In Figure \ref{f:sn1999bh_spec} we illustrate our two epochs of spectroscopy compared to \es\ at a comparable phase. In the top panel, we show our spectra around maximum light; \bh\ lacks hydrogen and exhibits \ion{Si}{2} $\lambda$6355, confirming its classification as a SN~Ia. In addition, there is strong \ion{Ti}{2} absorption, linking it to the \bg\ subclass. The ejecta velocity measured from the minimum blueshift of \ion{Si}{2} $\lambda$6355 is \about 6000 \kms, similar to that of \es. In the bottom panel, we show our spectra from roughly 3 weeks after maximum light. Again, we see striking similarities between the two objects. We note that the narrow feature at the wavelength of $\rm{H}{\alpha}$ in both spectra is from the host galaxy.

The host of \bh\ is classified by \cite{van-den-bergh02a} as an Sb galaxy. Given the narrow $\rm{H}{\alpha}$, [\ion {N}{2}], and \ion{Na}{1}~D features superimposed on our spectra of \bh, it is likely that \bh\ lies along the line-of-sight of a star-forming region in NGC 3435. However, it is not possible to determine whether \bh\ is actually associated with the star-forming region or if \bh\ is behind it. Unlike the case for \es, the host galaxy of \bh\ does not help constrain the stellar population associated with \bh.

\begin{deluxetable*}{cccccc}
\tablewidth{0pc}
\tablecaption{Comparison of Properties of Subluminous \bg-Like Objects \label{t:peculiar}}
\tablehead{
\colhead{SN} &
\colhead{$M_{B}$ (mag)} &
\colhead{$\Delta m_{15} (B)$ (mag)}  &
\colhead{$v_{\rm phot}$\tablenotemark{a} (\kms)}  &
\colhead{ $\mathcal{R}$(Si)} &
\colhead{Source}
}
\startdata
SN~1991bg  & $-16.60 \pm 0.03$ &  $1.93 \pm 0.10 $ &   10,100 & $0.62 \pm 0.05$ & \cite{taubenberger08a} \\
SN~2006bt  &  $-18.83 \pm 0.06$ &  $1.09 \pm 0.06 $ &   10,500 & $0.44 \pm 0.05$ & \cite{foley10b} \\
PTF~09dav &  $-15.33 \pm 0.08$ &  $1.87 \pm 0.06 $ &    6100  &$ 0.35 \pm 0.05$ & \cite{sullivan11a}  \\
PTF~10ops &  $-17.66 \pm 0.06$ &  $1.12 \pm 0.06 $ &   10,000 &$ 0.58 \pm 0.06$ & \cite{maguire11a} \\
SN~2002es &  $-17.78 \pm 0.12$ &  $1.28 \pm 0.04 $&   6000   & $0.55 \pm 0.05$ & This work
\enddata
\tablenotetext{a}{As measured by the minimum in the absorption feature attributed to \ion{Si}{2} $\lambda$6355.}
\end{deluxetable*}

\subsection{Rate}
Both \es\ and \bh\ were included in the LOSS SN rate study \citep{leaman11a,li11a,li11b} in the luminosity function (LF) subsample used to calculate the volumetric rate for different SN~Ia subtypes. The LF subsample is considered to represent a complete sample of 74 SNe~Ia within 80 Mpc. In \cite{li11b}, both objects are classified as \cx-like SNe. Within a fixed volume, \es-like objects should account for \about 2.5\% of SNe~Ia. It is worth noting that the reclassification of \es\ and \bh\ decreases the reported volumetric fraction of \cx-like objects from 5.0\% to 2.5\% as well. We caution that our rate calculation is limited by the small number of \es-like objects in the LF sample.

\subsection{Comparison to Other Peculiar \bg-Like SNe}
The spectroscopic subclass of \bg-like SNe has been shown to have their own form of Phillips relation (i.e., correlation between light-curve decline and luminosity), similar to what is found among normal SNe~Ia \citep{phillips99a,garnavich04a,taubenberger08a}. However, there are notable exceptions in the literature of SNe that share superficial similarities to \bg, but have unique properties not seen in the broader \bg\ subclass. Here we will compare and contrast \es\ with known peculiar \bg-like SNe. 

\cite{foley10b} found that SN~2006bt had broad, slowly evolving light optical curves ($\Delta m_{15}(B)= 1.09 \pm 0.06 ~\rm{mag}$) typical of a luminous event, but spectra more similar to those of a cooler, subluminous event. A spectrum of the host galaxy from the SDSS \citep{abazajian09a} shows no signs of emission lines, indicating no recent star formation. Coupled with the position of SN~2006bt in the halo of the galaxy, the authors concluded that the progenitor of SN~2006bt was likely a low-mass star. Assuming no host-galaxy extinction, the peak absolute magnitude of SN~2006bt was $M_{B} = -18.94 \pm 0.06$ mag, slightly less luminous than normal SNe~Ia, but not a significant outlier in the Phillips relation. SN~2006bt had a light-curve width similar to that of \es, but is significantly brighter. Unlike \es, the late-time decay in the light curves of SN~2006bt are consistent with \nic\ decay.  SN~2006bt had faster expansion velocities at maximum light compared to \es.

PTF has been a prolific discovery engine for unique transient events. Two subluminous SNe, in particular, may be related to \es : PTF~09dav \citep{sullivan11a} and PTF~10ops \citep{maguire11a}. Both events have maximum-light spectra that resemble those of \bg. PTF~09dav had slow expansion velocities of $\sim 6100$ \kms\ (similar to \es), while PTF~10ops had more typical velocities of $\sim 10,000$~\kms. Photometrically, both PTF~10ops and \es\  have broad light curves despite being subluminous. Both objects occupy a similar position off of the Phillips relation in Figure \ref{f:phillips}. PTF~09dav has a more typical \bg-like light-curve shape ($\Delta m_{15}(B) = 1.87 \pm 0.06$~mag), but  was $\sim 1.5$~mag fainter than typical subluminous SNe~Ia ($M_{B} = -15.5$~mag). However, \es\ is the only object where the optical light curves plummet after $t > +30~{\rm d}$.

In Table \ref{t:peculiar}, we summarize the photometric and spectroscopic properties of other peculiar \bg-like SNe in comparison to \es. While \es\ shares some characteristics with each of these SNe, no single previously published object is exactly like \es.

\subsection{Models}
Explaining the properties of \es\ within the confines of current models is challenging. Based on the star-formation history of the host galaxy of \es\ \citep{neill09a}, we expect the progenitor to be from a relatively old stellar population, indicating that it was an explosion of either a low-mass single star or a WD. In \S \ref{ss:cc} we examined the possibility that \es\ is the result of core collapse of a low-mass single star. In this section, we explore possible models to explain the observed properties of \es\ assuming a WD progenitor.

The ``.Ia" model has been proposed to explain rapidly evolving, subluminous thermonuclear events. They are the result of helium accretion onto a WD from a double-degenerate white dwarf AM Canum Venaticorum (AM~CVn) binary system \citep{bildsten07a,shen09a,shen10a} leading to a thermonuclear explosion of the accreted He envelope. These events are expected to have both a luminosity and a timescale that is 1/10th that of SNe~Ia. Calculations of observable properties by \cite{shen10a} find that the rise time of these objects is $ <  10$~d, with a fast decline after maximum. \es, on the other hand, has a rather broad light curve, with a rise that is likely longer than 10~d based on a first detection by LOSS of 9 days before the time of bolometric maximum light. The ejecta velocities are also expected to be $\sim 10^{4}$~\kms, almost a factor of 2 larger than what is observed in \es.

\cite{woosley11a} study a broader set of one-dimensional simulations with models involving a  sub-Chandrasekhar-mass carbon-oxygen (CO) WD undergoing helium accretion, including models that explode just the helium envelope and models that explode the entire star. The authors survey the parameter space of WD masses, accretion rates, and initial WD luminosities to produce model spectra and light curves for the resulting transient.

Models in which just the helium shell explode (either through detonation or deflagration) have rise times $< 10$~d and $\Delta m_{15}(B) > 2.0$~mag. The fast evolution of the $B$ band is caused by the small ejecta mass (depending on the size of the envelope) and the redistribution of flux from the optical to the NIR by IGEs in the ejecta \citep{kasen06a}. Detonations of ``cold" WDs ($L=0.01~{\rm L}_{\odot}$) produce spectra lacking IMEs, which are clearly seen in our spectra of \es. One of the more promising models from the set of helium-envelope explosions involved ``hot" WDs ($L=1~{\rm L}_{\odot}$). These explosions produce a significant amount of IMEs, resulting in spectra resembling \bg\ with ejecta velocities \about 9000 \kms. These models produce a very small amount of \nic\ (\about$10^{-4} ~{\rm M}_{\odot}$) and are instead powered by $^{48}{\rm Cr}$, producing a light curve with a 3 d rise time and peak absolute magnitude of $-13$. A light curve powered by $^{48}{\rm Cr}$ and subsequent decay to $^{48}{\rm V}$ (half life $\tau=16~\rm{d}$) is an appealing explanation for the rapid fading of \es\ at $t > +30$~d; however, these models evolve too fast and have ejecta velocities that are inconsistent with what we observe. 

Deflagrations within the helium envelope led to incomplete burning, producing lower ejecta velocities (\about 4000 \kms) and broader light curves caused by the increased diffusion time in comparison to the detonation model of comparable brightness. The light curves of these models are also powered by the decay of $^{48}{\rm Cr}$, producing faint transients ($M_{B} \approx -14$~mag). However, the evolution of these light curves is still much too fast compared to \es. 

Outcomes from exploding the entire star generally depended on the initial luminosity of the WD. ``Cold" WDs required a larger accreted mass to ignite the helium envelope and drive an explosion of the entire star. As a result, the ``cold" models contained an outer envelope of IGEs (synthesized from helium) which acted as a heating source to ionize IMEs and reduce their opacity. Model spectra of these events lack \ion{Si}{2}, \ion{S}{2}, and \ion{Ca}{2}, all of which are present in \es. The ``hot" models, on the other hand, explode with a smaller accreted envelope, allowing IMEs to appear in the spectra. In fact, the synthetic spectra for these objects look remarkably like those of normal SNe~Ia and even produce a width-luminosity relation, similar to the Phillips relation, but with a different slope. However, the minimum expansion velocities seen in these models are \about 11,000 \kms, much higher than what is observed for \es.

\cite{pakmor10a,pakmor11a} find in simulations that the merger of nearly equal mass WDs leads to underluminous explosions similar to \bg. Their model follows the evolution and subsequent explosion of two WDs of equal mass ($0.89~{\rm M}_{\odot}$). They find explosions that have roughly the same kinetic energy as a normal SN~Ia (\about$10^{51}~{\rm erg~s^{-1}}$), but lower velocities due to the larger ejecta mass. Less \nic\  ($0.1~{\rm M}_{\odot}$) is synthesized due to lower densities in the final merged object, resulting in a subluminous event. Synthetic light curves of their models are similar to those of \bg, but broader, with $\Delta m_{15}(B) = 1.4$--1.7 mag. Synthetic spectra show strong titanium, as well as the presence of IMEs such as \ion{Si}{2} and \ion{O}{1} at velocities lower than typically seen in \bg-like objects. All of these match characteristics seen in \es. However, their models do not predict the fast drop in flux at $t > +30$ d, which is a key characteristic that makes \es\ a unique object.


\section{Conclusions}\label{s:conclusions}
\es\ is a peculiar, subluminous SN~Ia with a unique combination of observables. At maximum light, spectra of \es\ are similar to the subluminous \bg, indicating a cool photosphere, but with ejecta velocity of \about 6000 \kms. Such slow velocities are more characteristic of \cx-like objects. While also subluminous, \cx\ had a maximum-light spectrum resembling that of SN~1991T, which is characteristic of a hot photosphere.

Photometrically, \es\ has a broad light curve ($\Delta m_{15}(B) = 1.28~\rm{mag}$), despite being subluminous with a peak absolute magnitude of $M_{B} = -17.78$~mag. \es\ is a $5\sigma$ outlier in the Phillips relation \citep{phillips93a,phillips99a} used to calibrate the light-curve width vs. luminosity relationship for SNe~Ia. The $R$- and $I$-band light curves are broad and lack the shoulder typically seen in SNe~Ia.

Quantitative measurements of spectral features such as the silicon ratio \citep[$\mathcal{R}({\rm Si});$][]{nugent95b} and the \ion{Si}{2} $\lambda$6355 velocity gradient are similar to those of \bg-like objects. However, \es\ is an outlier in the usual relationships, which show strong correlations between these spectral measurements and light-curve parameters \citep{benetti05a,branch06a}.

From Arnett's law, we estimate a synthesized radioactive nickel mass of $0.17~{\rm M}_{\odot}$ required to power the light curve. However, the bolometric light curve shows an unexpected drop in luminosity at $t > +30$~d. We are unable to fit the bolometric light curve with a toy model of the radiated luminosity that accounts for a decrease in the $\gamma$-ray trapping function as the ejecta expand homologously. If \es\ is a thermonuclear event, then the ejecta became optically thin to $\gamma$-rays in an unexpectedly dramatic fashion. Similarly, we are unable to explain the rapid decay by invoking dust formation or the infrared catastrophe \citep{axelrod80a}. Alternatively, \es\ could be powered by some other mechanism not yet understood.

\es\ exploded in the outskirts of the early-type S0 galaxy UGC 2708. SED fitting from \cite{neill09a} and emission-line diagnostics indicate that UGC 2708 is likely a LINER galaxy with no current star formation. This points to a old star, likely a WD, as the progenitor to \es.

Finding a published model that matches the peculiar collection of observables found in \es\ is particularly challenging. We have not found a convincing match by comparing \es\ to \citet{woosley11a} models of WDs undergoing helium accretion from a companion star. Models of the merging and subsequent detonation of two equal-mass WDs from \cite{pakmor10a,pakmor11a} are promising, but do not reproduce the drop in luminosity one month after maximum light.

Looking through the existing LOSS photometry \citep{ganeshalingam10a} and the Berkeley Supernova Ia Program spectral database \citep[BSNIP; ][]{silverman12a}, we identify \bh\ as a probable \es-like event. Spectra and photometry of the object, while limited, match many of the observed characteristics of \es. Both objects are included in the LOSS SN rate studies \citep{leaman11a,li11a,li11b}, allowing us to estimate that roughly 3\% of SNe~Ia should be \es-like SNe within a fixed volume.

Ongoing surveys for transient objects, such as PTF and Pan-STARRS, are likely to find more objects like \es. PTF has already published results for two peculiar subluminous objects which share some similarities to \es, but also have distinct differences. We expect that future datasets will shed light on understanding the bizarre nature of these subluminous objects.

\section*{Acknowledgments}

We thank the Lick Observatory staff for their assistance with the operation of KAIT. We also thank B. Barris, P. Berlind, S. Jha, M. Papenkova, and B. Swift for their help with some of the observations. M.G. acknowledges useful and insightful conversations with  D. Kasen and I. Kleiser regarding theoretical models of SNe~Ia. M.G. is very grateful to S. Blondin for providing unpublished spectra of \es\ and \bh\ that will be included in an upcoming CfA SN~Ia spectroscopy paper. P.A.M. thanks G. Williams and P. Watje for late-epoch observations of SN~2002es.

The research of A.V.F.'s supernova group at UC Berkeley has been generously supported by the US National Science Foundation (NSF; most recently through grants AST--0607485 and AST--0908886), the TABASGO Foundation, US Department of Energy SciDAC grant DE-FC02-06ER41453, and US Department of Energy grant DE-FG02-08ER41563.  KAIT and its ongoing operation were made possible by donations from Sun Microsystems, Inc., the Hewlett-Packard Company, AutoScope Corporation, Lick Observatory, the NSF, the University of California, the Sylvia \& Jim Katzman Foundation, the Richard and Rhoda Goldman Fund, and the TABASGO Foundation. Supernova research at Harvard is supported in part by NSF grant AST--0907903. We made use of the NASA/IPAC Extragalactic Database (NED), which is operated by the Jet Propulsion Laboratory, California Institute of Technology, under contract with NASA.

\bibliographystyle{apj}


\end{document}